\documentclass{preparation}

\usepackage{epsfig}
\usepackage{color}
\usepackage{bm}
\usepackage{graphicx}
\usepackage{longtable}
\usepackage{amssymb}
\usepackage{rotating}
\usepackage{lineno}
\usepackage{float}
\usepackage{booktabs}
\usepackage{amsmath}
\usepackage{capt-of}

\title{A Wandering 35,000-Solar-Mass Black Hole Fed by a Gravitational Wake}

\author{Xin Li$^{1}$, Yong Shi$^{1\,\ast}$, Fuyan Bian$^{2,3~\ast}$, Junfeng Wang$^{4~\ast}$, Shude Mao$^{1}$, Qiusheng Gu$^{5}$, Yifei Jin$^{1}$, Yanmei Chen$^{5}$, Zhiyuan Zheng$^{6}$, Qinwei Yuan$^{5}$, Xiaoling Yu$^{7}$}

\begin{document}

\maketitle
\let\thefootnote\relax\footnote{
\begin{affiliations}
   \item Department of Astronomy, School of Science, Westlake University, Hangzhou, Zhejiang 310030, China.
  \item European Southern Observatory, Alonso de Córdova 3107, Casilla 19001, Vitacura, Santiago 19, Chile.
  \item Chinese Academy of Sciences South America Center for Astronomy, National Astronomical Observatories, CAS, Beijing 100101, China.
  \item Department of Astronomy and Institute of Theoretical Physics and Astrophysics, Xiamen University, Xiamen, Fujian 361005,  China.
  \item School of Astronomy and Space Science, Nanjing University, Nanjing 210093, China.
  \item Research Center of Astronomy, Qinghai University, Xining, 810016, China.
  \item College of Physics and Electronic Engineering, Qujing Normal University, Qujing, Yunnan 655011, China.\\
  $\ast$ Corresponding authors. E-mails: shiyong@westlake.edu.cn; fuyan.bian@eso.org; jfwang@xmu.edu.cn.
  
\end{affiliations}
}
\vspace{-3.5mm}

\begin{abstract}
Intermediate-mass black holes are widely considered to be the seeds of supermassive black holes \cite{Volonteri10, Greene20}, a substantial population of which is expected to remain displaced from galactic nuclei owing to hierarchical galaxy assembly and inefficient dynamical friction \cite{Bellovary19, Ricarte21, Beckmann23}. While several fueling channels can sustain central black holes, those pathways are largely inaccessible to off-nuclear black holes, leaving their fuel supply uncertain. As these wandering black holes move through the interstellar medium of their host galaxies, theory predicts that they can capture gas from the dense wake produced by gravitational focusing \cite{Edgar04, Ogata24}. However, direct observational evidence for this process has remained elusive. Here we report evidence for a wandering intermediate-mass black hole of 35,000 solar mass accreting through such a gravitational wake. Its black-hole nature is supported by broad-line emission, a compact continuum counterpart, long-term optical variability, and a power-law-like spectral energy distribution. Multi-epoch spectroscopy reveals three distinct gas components: a blueshifted, low-density upstream flow; a redshifted, dense downstream wake; and optically thick absorbers well within the capture radius that drive rapidly changing-look variability in the broad-line emission. This discovery establishes a previously unobserved channel for the growth of wandering intermediate-mass black holes. 
\end{abstract}

\begin{figure}
\centering
\includegraphics[width=0.68\textwidth]{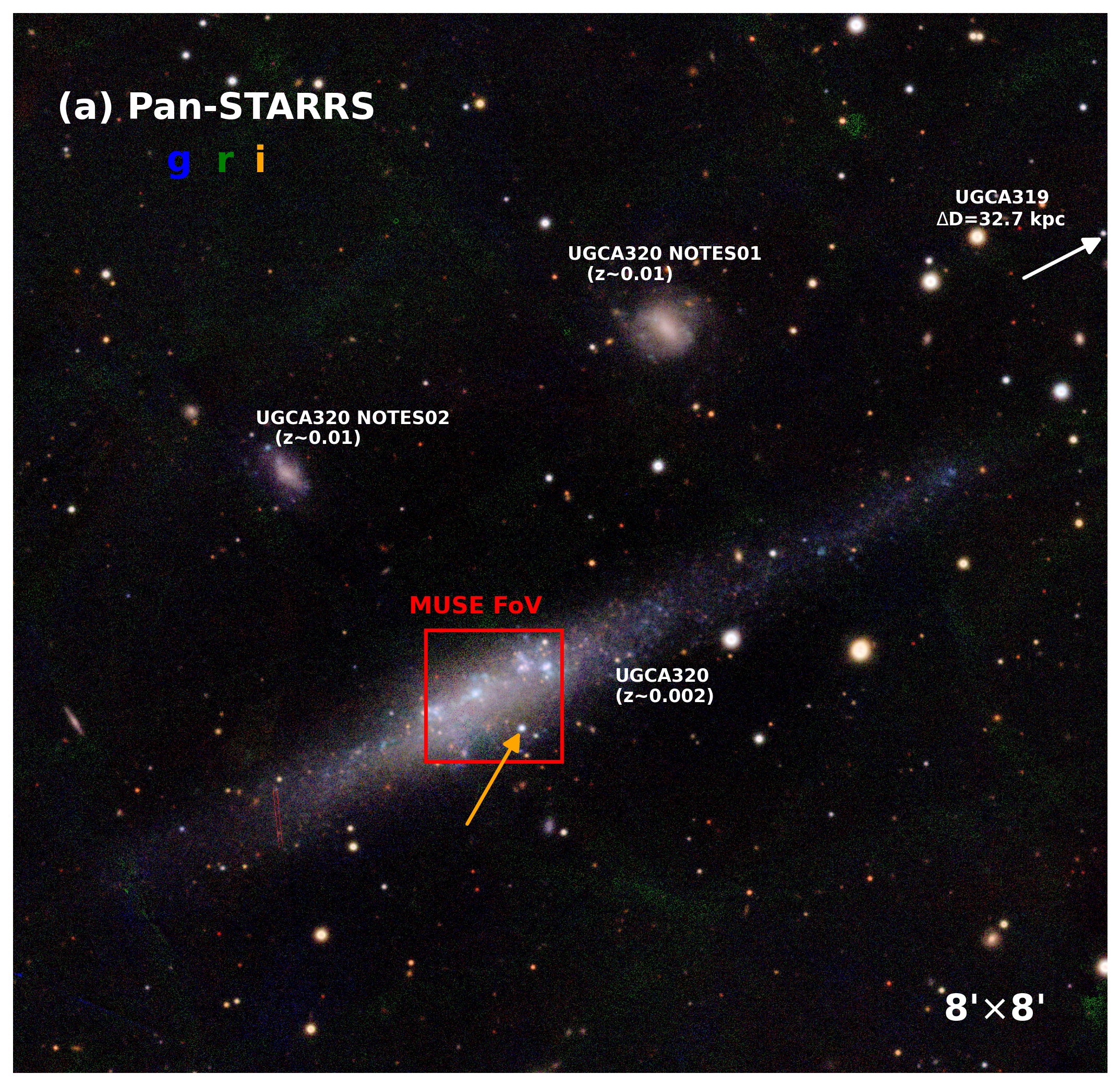}\\
\includegraphics[width=0.43\textwidth]{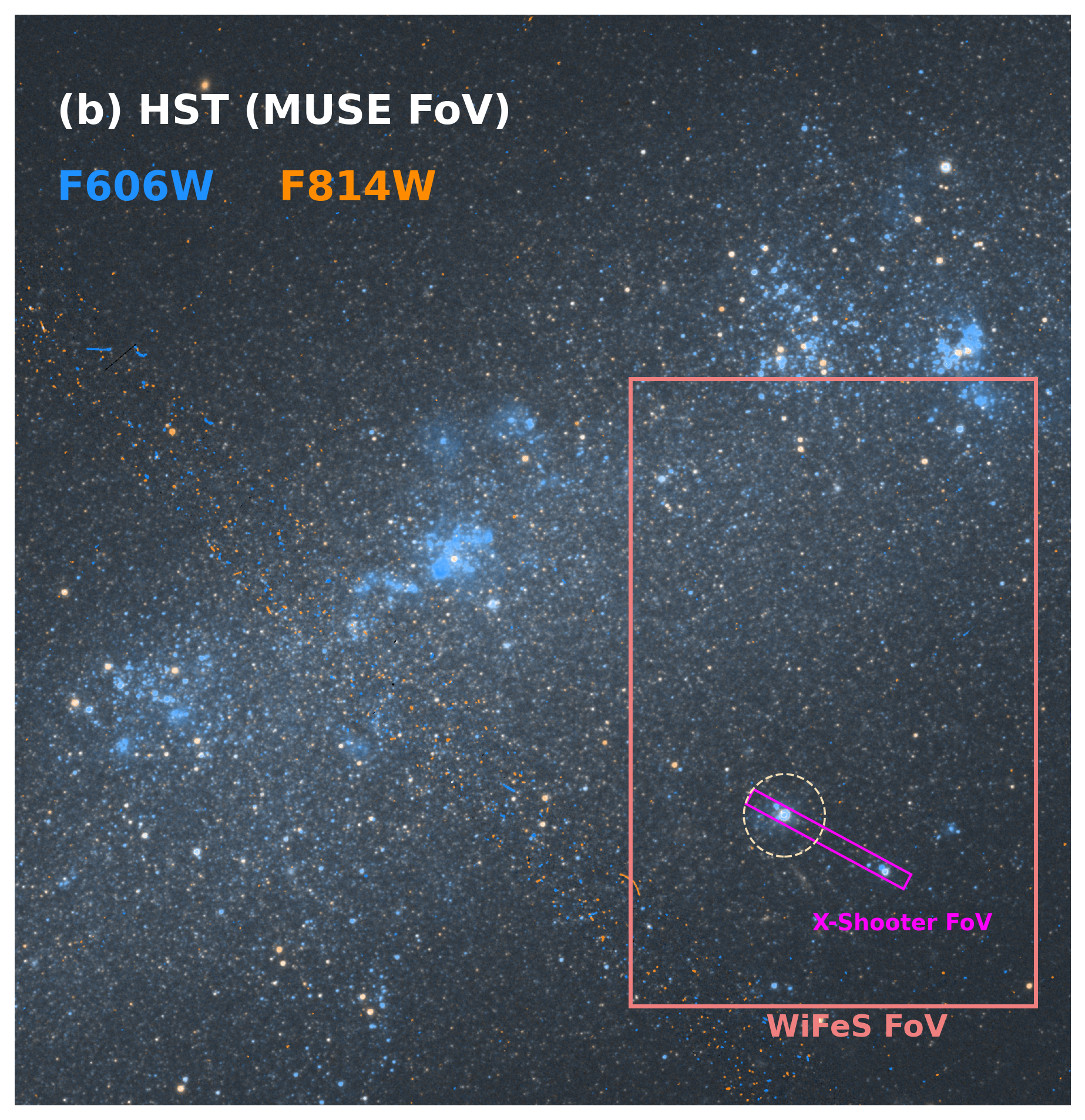}
\includegraphics[width=0.25\textwidth]{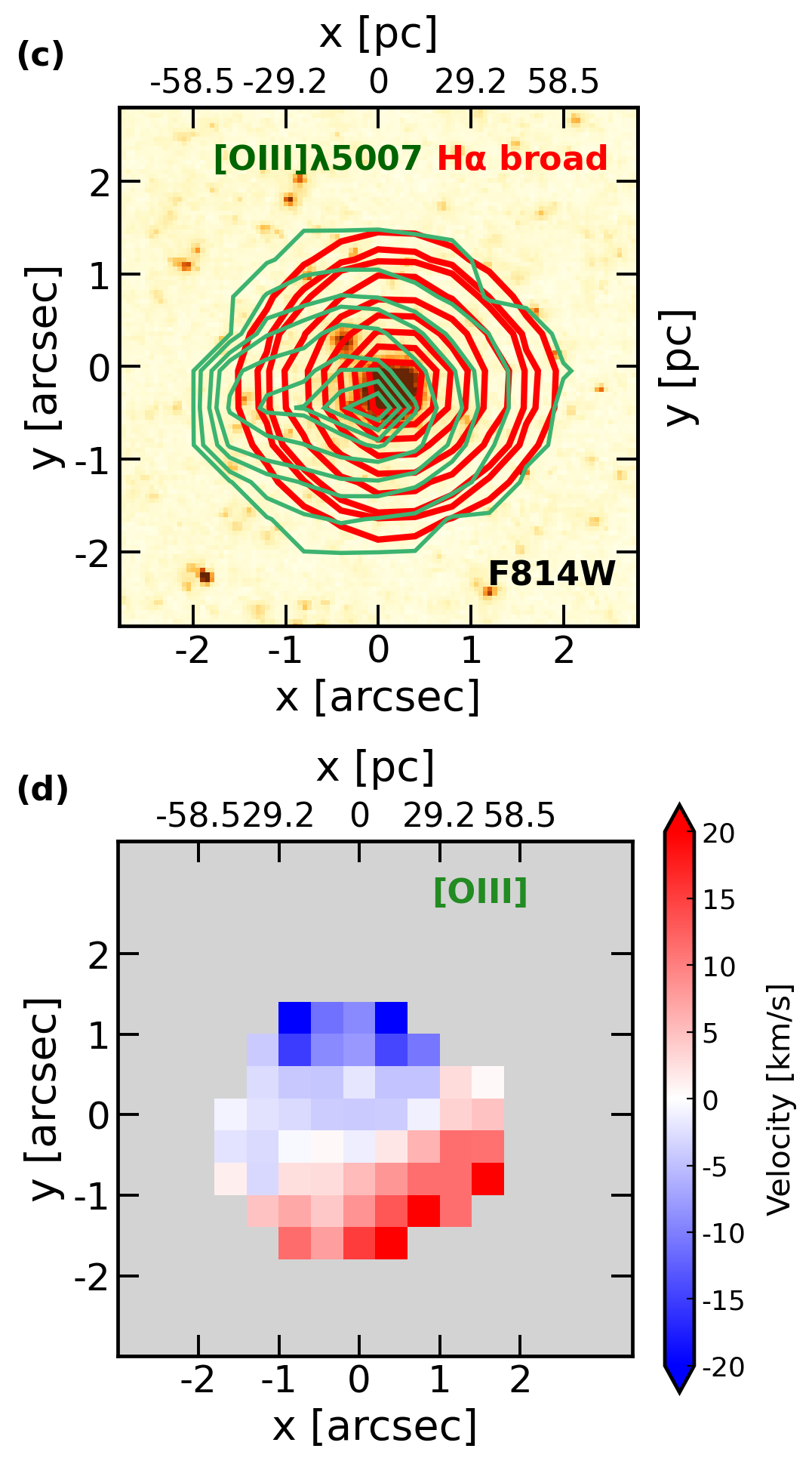}
\caption{\label{fig:img} \textbf{Overview of UGCA 320 and UGCA320-IMBH.} {\bf a,} Pan-STARRS false-color image of UGCA 320, with the MUSE field of view (FOV) outlined in red. The white arrow indicates the direction of the paired galaxy UGCA 319, while the yellow arrow marks the position of the discovered wandering intermediate-mass black hole (IMBH), UGCA320-IMBH. {\bf b,} \textit{HST} false-color image of UGCA 320, centered on the MUSE FOV and overlaid with the WiFeS (tomato) and X-Shooter (magenta) FOVs. The yellow dashed circle marks a 2$''$-radius region centered on UGCA320-IMBH. {\bf c,} Zoomed-in \textit{HST} F814W image centered on UGCA320-IMBH, overlaid with contours of the broad H$\alpha$ flux (red) and [O\,{\sc iii}]$\lambda5007$ flux (green) derived from the MUSE data. {\bf d,} [O\,{\sc iii}]$\lambda$5007 velocity map of UGCA320-IMBH region, after subtraction of the local systemic velocity.}
\end{figure}

The host galaxy UGCA 320 (13:03:16.74, $-$17:25:22.9), also known as DDO 161, is a nearby edge-on dwarf irregular galaxy located at a distance of $D = 6.03^{+0.29}_{-0.21}\,\mathrm{Mpc}$ (ref-\citenum{Karachentsev17}). It has a stellar mass of $M_{\ast}=1.3\times10^8~M_{\odot}$ (ref-\citenum{Cook14}), H\,{\sc i} mass of $M_{\rm HI}=9.7\times10^8~M_{\odot}$ (ref-\citenum{deBlok24}), and a star-formation rate of $\log {\rm SFR(UV)}=-1.4~M_{\odot}\,{\rm yr^{-1}}$ (ref-\citenum{LX26}). UGCA 320 resides in a field environment within the Local Volume and forms a pair with the nearby dwarf galaxy UGCA 319 in the north. Recent studies have further shown that it hosts an unusually rich population of satellite galaxies \cite{LJ26}. Optical imaging reveals a faint distortion toward the companion (Fig.~\ref{fig:img}a) indicative of weak interaction \cite{Alabi25, Zabel26}. In addition, H{\sc i} observations have revealed a galaxy-scale outflow extending along the minor axis\cite{Zabel26}. We observed this source with the Multi Unit Spectroscopic Explorer (MUSE) \cite{Bacon10} on the Very Large Telescope (VLT) in 2021 (ref-\citenum{LX26}), and subsequently carried out follow-up \textit{Chandra} \cite{Garmire03} X-ray observations, along with additional optical spectroscopy using X-shooter \cite{van01} on the VLT (2025) and the Wide Field Spectrograph (WiFeS) \cite{Dopita10} on the Australian National University (ANU) 2.3-m telescope (2025 and 2026). These data were complemented by archival ground-based and \textit{Hubble Space Telescope} (\textit{HST}) \cite{Sirianni05} imaging.

The newly identified off-nuclear object (13:03:15.8902, $-$17:25:36.591) is associated with the host galaxy through their consistent spectroscopic redshifts. The \textit{HST} false-color image shows that it lies outside the main star-forming disc of the galaxy, where contamination from host emission is minimal (Fig.~\ref{fig:img}b). Multiple independent observations consistently identify the source as an accreting intermediate-mass black hole (IMBH) as detailed below.

First, VLT/MUSE observations obtained in 2021 reveal broad Balmer emissions, the spectroscopic signature of a type-1 accreting massive black hole. The broad H$\alpha$ component has a full width at half maximum (FWHM) of 854$\pm$6.02 km/s. Under the black hole scenario, it corresponds to a black hole mass of $3.5\times10^4~M_{\odot}$ derived based on ref-\citenum{Reines13}. 

Second, the peak of broad H$\alpha$ emission identified in the MUSE observations is spatially coincident with an unresolved continuum counterpart in the \textit{HST} ACS F814W image, at a spatial resolution of 2.8 pc (Fig.~\ref{fig:img}c). Such a compact continuum morphology rules out Balmer-dominated supernova remnants, whose emission-line spectra can resemble those of type-1 active galactic nuclei (AGN) but lack continuum emission \cite{Smith91, Li21}. The object also exhibits a relatively low [S\,{\sc ii}]/H$\alpha$ ratio compared with the host galaxy, inconsistent with shock-dominated supernova remnants (see Extended Data Fig.~\ref{fig_bpt} in Methods). The Baldwin–Phillips–Terlevich (BPT) diagnostic diagrams \cite{Baldwin1981} (Extended Data Fig.~\ref{fig_bpt}) place UGCA320-IMBH within the star-forming locus, exhibiting even lower [O\,{\sc iii}]/H$\beta$, [N\,{\sc ii}]/H$\alpha$, and [S\,{\sc ii}]/H$\alpha$ ratios than the host galaxy. The suppressed line ratios can be due to low ionization, low metallicity, and the attenuation of the UV photons \cite{Inayoshi25b, Maiolino26, Ji26}.

Third, the object displays AGN-like stochastic optical variability over the past 14 years (Fig.~\ref{fig:change}a), and \textit{Gaia} Data Release 3 epoch observations also identified it as a variable source\cite{GaiaColl23}. After a brighter phase more than a decade ago, it faded to a minimum luminosity around 2019 before gradually brightening again.  The observed variability amplitude of $\sim$ 0.15 mag is also substantially smaller than that typically observed in luminous blue variables (LBVs) \cite{Humphreys14, Spejcher25}, which exhibit strong Balmer emission in quiescence and p-Cygni absorption during eruptive phases \cite{Smith11, Humphreys14, Humphreys17}.

\begin{figure}
\centering
\includegraphics[width=0.99\textwidth]{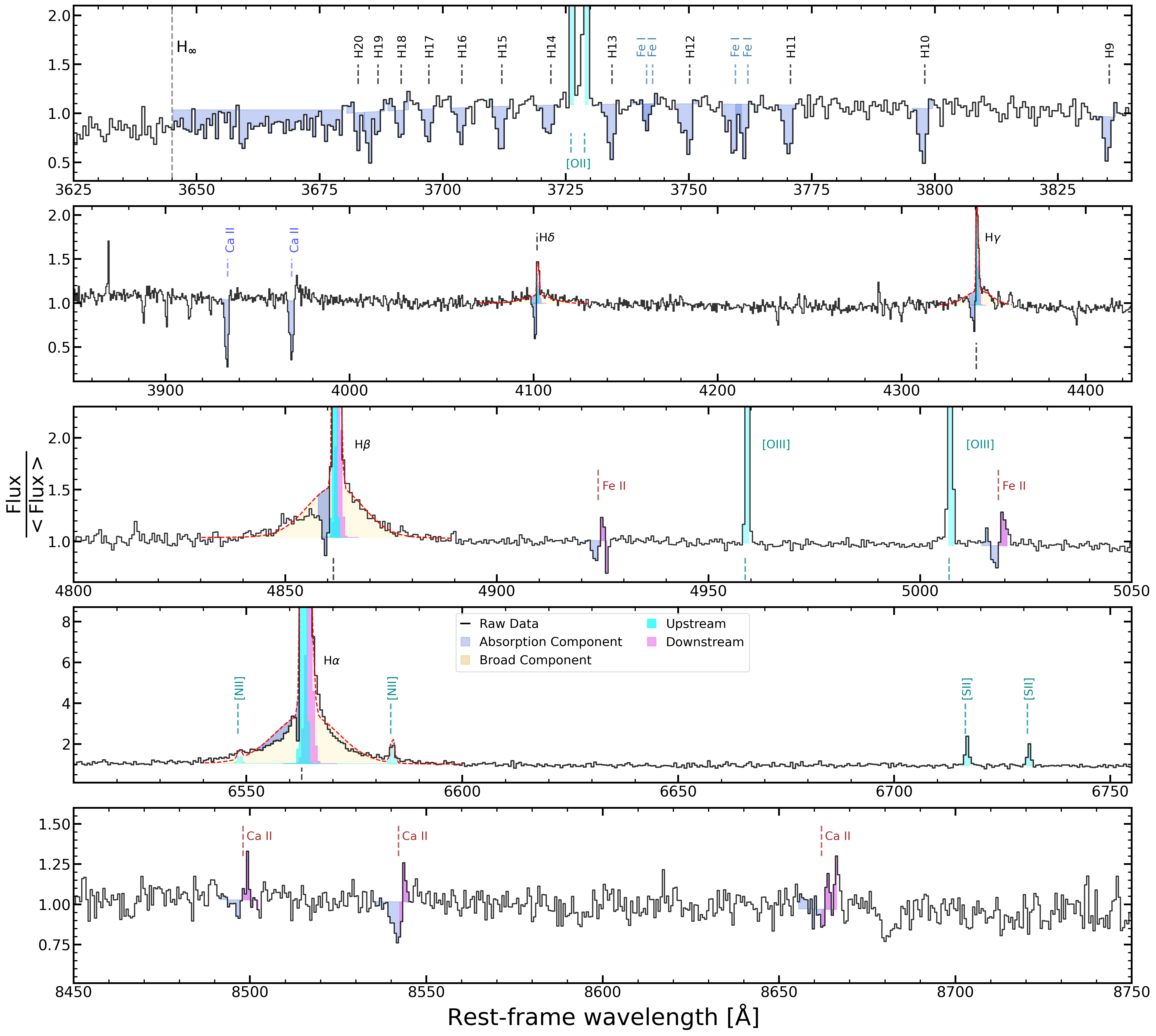}
\caption{\label{fig:spec} \textbf{X-Shooter spectrum of the UGCA320-IMBH.} The spectrum in each panel is normalized by its median flux. Dashed vertical lines mark the rest-frame line centers. Spectral components sharing the same kinematics are shown in the same color: a blueshifted absorption component produced by the accreted gas clump (blue), a blueshifted narrow emission component tracing ambient upstream flow (cyan), a redshifted narrow emission component tracing dense downstream wake (magenta), and a broad Balmer-line component associated with the active galactic nucleus (yellow). The red dashed curves overlapped on the H$\delta$, H$\gamma$, H$\beta$, and H$\alpha$ show the Gaussian models incorporating all emission components.}
\vspace{-4mm}
\end{figure}

Fourth, the UV-to-optical spectral energy distribution (SED) can be well described by an AGN-power-law continuum partially attenuated by optically thick gas ($\tau_{\rm b}\gg1$ at $\lambda\approx3646$\,\AA) together with modest dust extinction (Fig.~\ref{fig:change}b). In contrast, stellar atmosphere models cannot simultaneously reproduce both the ultraviolet and optical continua: hot-star models are too steep in the optical, whereas warm-star models are too weak in the ultraviolet (see Extended Data Fig.~\ref{fig_star}). The bolometric luminosity $L_{\rm bol}$ inferred from the SED also exceeds the empirical upper luminosity limit of known massive stars in the Milky Way (MW) \cite{Humphreys13, Humphreys14}, making it too bright to be a single star. Even when modeled as a binary stellar system, both components exceed the empirical luminosity boundary at their corresponding effective temperatures (Extended Data Fig.~\ref{fig_star}). Furthermore, the X-Shooter spectrum shows no detectable 4000\,\AA\ break (Fig.~\ref{fig:spec}), contrary to the prediction of stellar atmosphere models. In addition, the X-Shooter spectrum shows high-order Balmer absorption lines that exhibit nearly identical equivalent widths, indicating that the absorption has reached the saturated regime. Such strong absorption cannot be produced in a stellar atmosphere\cite{Inayoshi25, Ji25, Lin26}. Consistent with these findings, young massive binary systems are typically embedded in star-forming regions, whereas the \textit{HST} image shows that the optical emission is dominated by the central point source ($\sim$90\% relative to the $\sim1''$-radius SED aperture), with only a few nearby compact blue sources, unlike the numerous compact blue knots that characterize the star-forming regions in the main disc.

\begin{figure}
\centering
\includegraphics[width=0.8\textwidth]{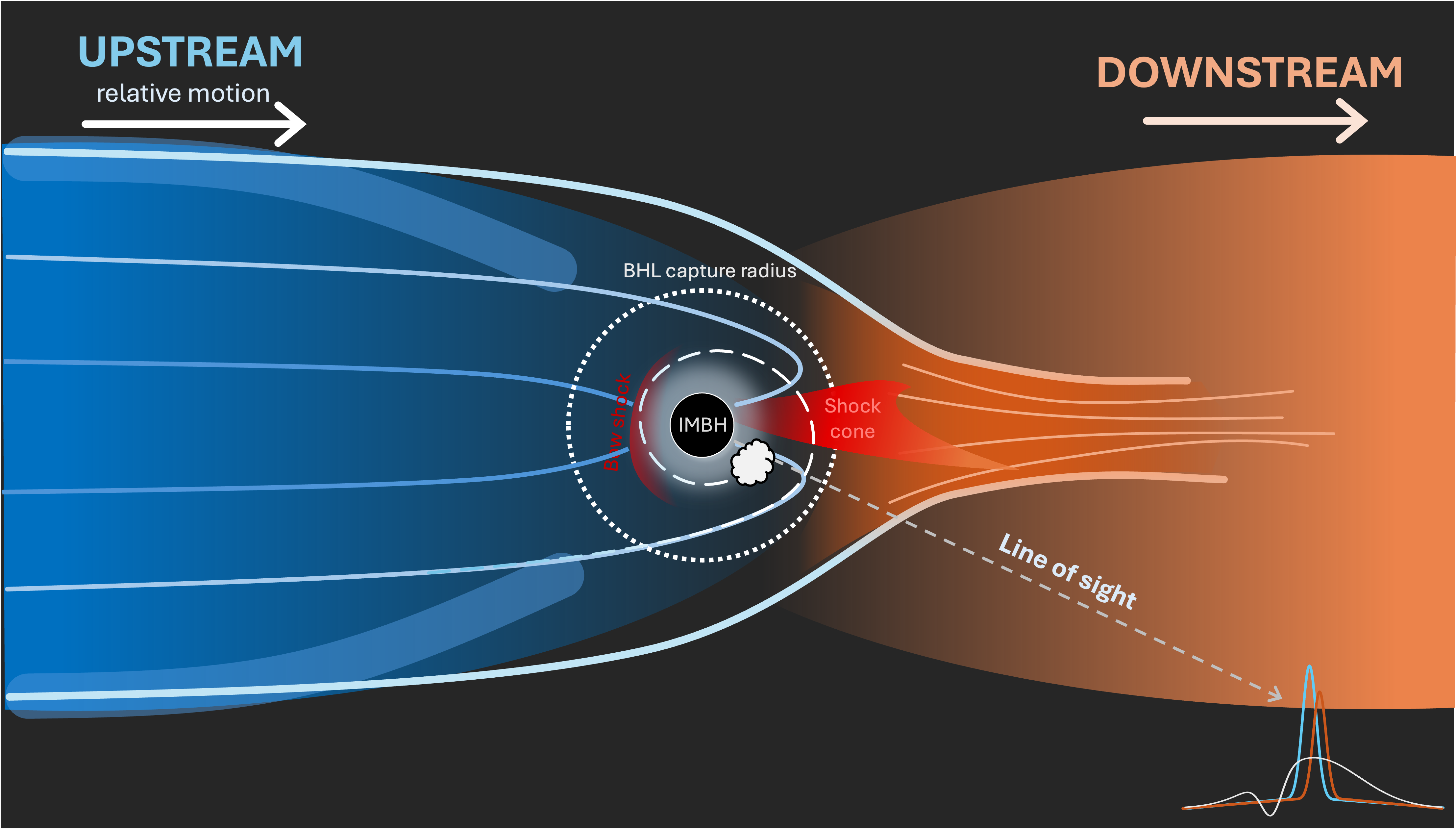}
\caption{\label{fig:BHL} \textbf{Schematic illustration of Bondi–Hoyle–Lyttleton accretion onto a wandering intermediate-mass black hole.} In the rest frame of the intermediate-mass black hole (IMBH), the ambient gas flows from the upstream region (blue, left) towards the downstream region (orange, right), as indicated by the solid streamlines. Gravitational focusing deflects and compresses gas within the approximate Bondi–Hoyle–Lyttleton (BHL) capture radius (white dotted circle) towards the IMBH. The converging flow produces an upstream bow shock and a compressed, dense accretion wake downstream with a shock cone behind the black hole. The captured gas circulates around the black hole before eventually being accreted (white dashed curve), and partially covers the central emitting region along the observer’s line of sight (gray dashed line). The line profiles at the lower right schematically illustrate the blueshifted upstream and redshifted downstream narrow-line components (blue and orange, respectively), together with broad Balmer emission affected by blueshifted absorption (white). }
\vspace{-4mm}
\end{figure}

In contrast to black holes at galactic nuclei, which can access diverse fueling channels—including galaxy mergers, tidal interactions, cloud–cloud collisions, and gas cooling—to funnel gas into the central potential, such pathways are largely unavailable to off-nuclear black holes. One plausible accretion channel for a wandering black hole is through the gravitational wake it generates while moving through the interstellar medium, known as Bondi–Hoyle–Lyttleton accretion (BHL)\cite{Edgar04, Ogata24}. In this scenario, the surrounding gas is expected to develop an asymmetric flow structure, consisting of a low-density upstream component from the ambient interstellar medium ahead of the black hole, a dense downstream wake formed by gravitational focusing behind it, and an inner accretion flow formed as the black hole captures gas from the wake. A schematic illustration of this configuration is shown in Fig.~\ref{fig:BHL}. Our spectroscopic observations reveal all three key components predicted by the BHL framework. First, our MUSE observations reveal a clear [O\,{\sc iii}]$\lambda5007$ velocity gradient despite the $\sim$28 pc spatial resolution of the data (Fig.~\ref{fig:img}d), indicating blueshifted and redshifted gas components surrounding the IMBH. Further evidence for multiple components with different velocities and densities comes from the X-shooter/VLT long-slit spectrum toward the black hole, stacked within a 1$''$ aperture. 

{$\bullet$ Upstream gas with low  density:} In the X-shooter spectrum (Fig.~\ref{fig:spec}), forbidden emission lines such as [O\,{\sc iii}] and [S\,{\sc ii}] exhibit a blueshifted velocity of $-27~\mathrm{km\,s^{-1}}$ relative to the local systemic velocity. The detected [S\,{\sc ii}] doublet ratio\cite{Proxauf14} implies an electron density of $\sim 40~\mathrm{cm^{-3}}$. This density is consistent with typical interstellar medium conditions, suggesting that this blueshifted component traces the low-density upstream gas flowing toward the black hole.

{$\bullet$ Downstream wake with high density:} In the X-shooter spectrum (Fig.~\ref{fig:spec}), we identify a second set of narrow emission lines at a redshifted velocity of $+27~\mathrm{km/s}$, marked by a series of permitted Fe\,{\sc ii} and Ca\,{\sc ii} transitions. The presence of these lines indicates that the gas is partially ionized \cite{Baldwin04, Martinez21}. Although accurately constraining its density is challenging,  the coexistence of strong permitted Fe\,{\sc ii} and Ca\,{\sc ii} emission nevertheless requires gas densities orders of magnitude higher than the ambient interstellar medium, likely exceeding $10^6~{\rm cm^{-3}}$ (ref-\citenum{Matsuoka07, Martinez15}). This dense, redshifted component is naturally interpreted as the gravitationally focused downstream wake predicted by the BHL accretion scenario.

{$\bullet$ Dense clumps tracing the accretion flow:} In addition to the two velocity components traced by emission lines, we identify a third component through a series of absorption features, including the Balmer series from H$\alpha$ down to the Balmer break, as well as Fe\,{\sc i}, Fe\,{\sc ii}, and Ca\,{\sc ii}. These absorption lines are all centered at a similar blueshifted velocity of $\sim -50~\mathrm{km/s}$. The inferred hydrogen column density should exceed $5\times10^{22}\,\mathrm{cm^{-2}}$ in order to reproduce both the Balmer break and the Balmer equivalent widths (Extended Data Fig.~\ref{fig_cloudy}). Close in time to our X-shooter observations, we carried out deep \textit{Chandra} X-ray observations but detected no significant signal; this non-detection independently implies a column density of $N_{\rm H}\gtrsim10^{23}~\mathrm{cm^{-2}}$.

Multi-epoch spectroscopic monitoring further reveals changing-look behavior in the broad H$\alpha$ and H$\beta$ emission lines over a timescale of only a few years (Fig.~\ref{fig:change}): both broad components were prominent in April 2021, nearly disappeared in April and June 2025, partially reappeared in July 2025, and faded again by April 2026. We can place an order-of-magnitude constraint on the location of the obscuring absorber. Assuming its orbital velocity is comparable to the observed line-of-sight velocity ($\sim50$ km s$^{-1}$), the corresponding Keplerian radius for a black hole mass of $3.5\times10^4~M_\odot$ is $\sim0.06$ pc. This radius would be smaller if the absorber has a substantial velocity component perpendicular to the line of sight. The inferred location lies well within the BHL capture radius of $\sim0.3$ pc, assuming a sound speed of 10~km\,s$^{-1}$ and a black-hole relative velocity of 30~km\,s$^{-1}$, indicating that the absorbing clumps are embedded in the accretion flow toward the central black hole. The rapid variability of the broad-line component further favors a smaller absorber radius. For example, if the obscuring cloud has a characteristic size of 100 AU, the required crossing speed would be as large as $\sim500$ km s$^{-1}$ for a crossing timescale of 1 year.



Our discovery provides the observational evidence that wandering intermediate-mass black holes can actively accrete through gravitational wakes. As a substantial population of seed black holes is expected to spend extended periods away from galactic centers, this mobile accretion channel may represent an important pathway for the early growth of intermediate-mass black holes before they sink into galactic nuclei and contribute to the assembly of supermassive black holes.

 \clearpage

\begin{figure}
\begin{center}
\includegraphics[width=0.516\textwidth]{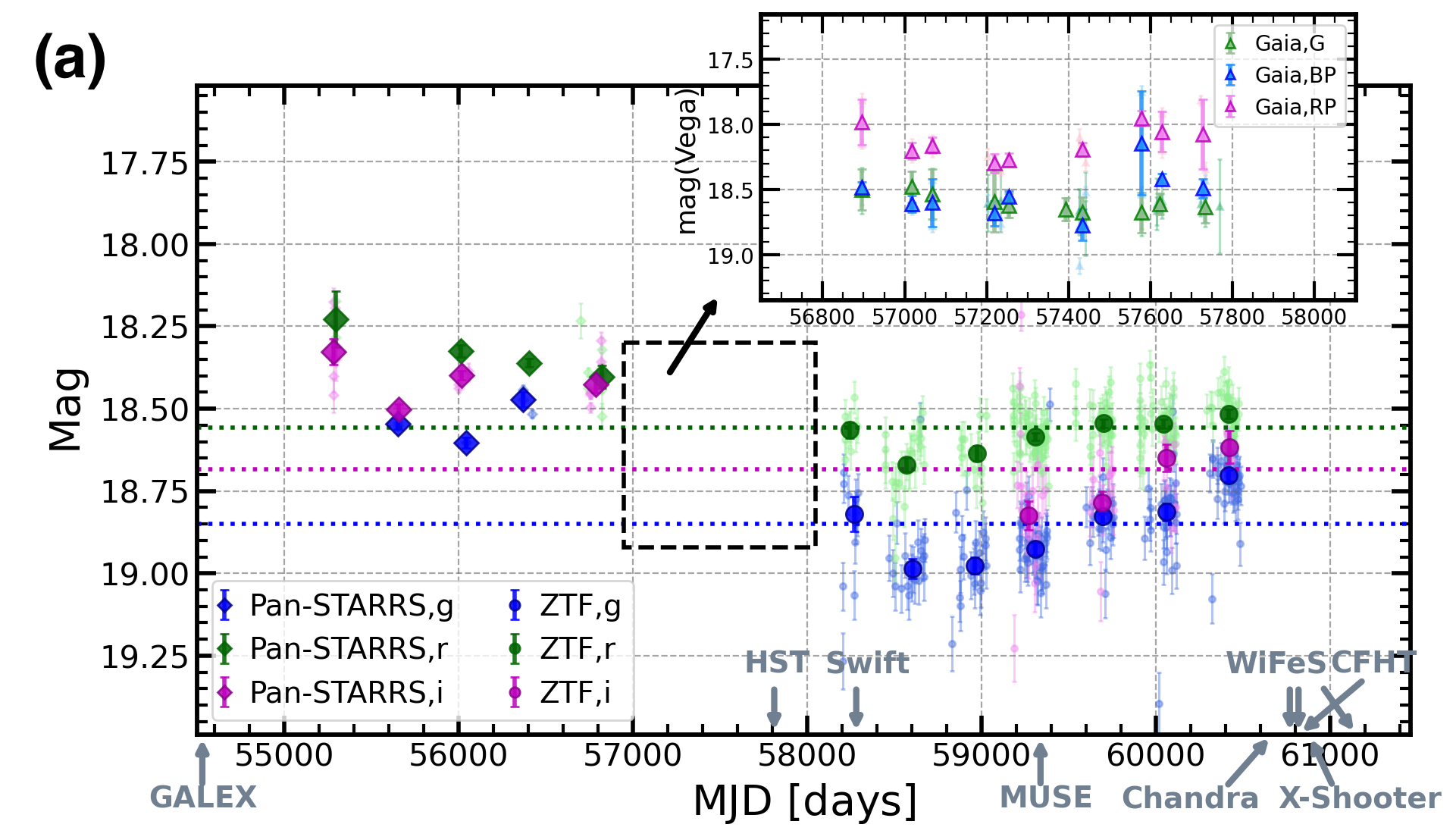}
\includegraphics[width=0.478\textwidth]{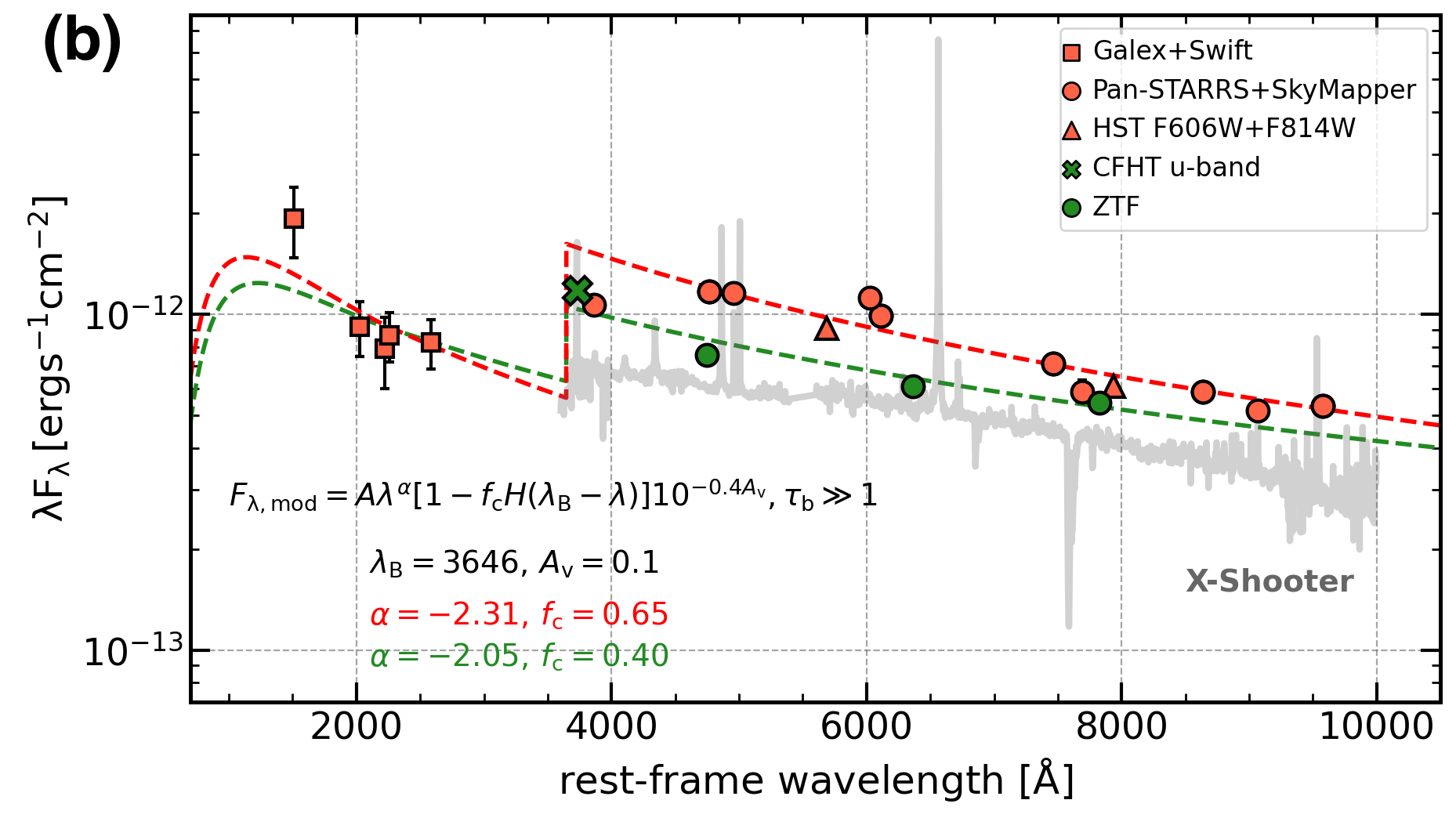}\\
\vspace{0.4cm}
\centering{\includegraphics[width=0.9\textwidth]{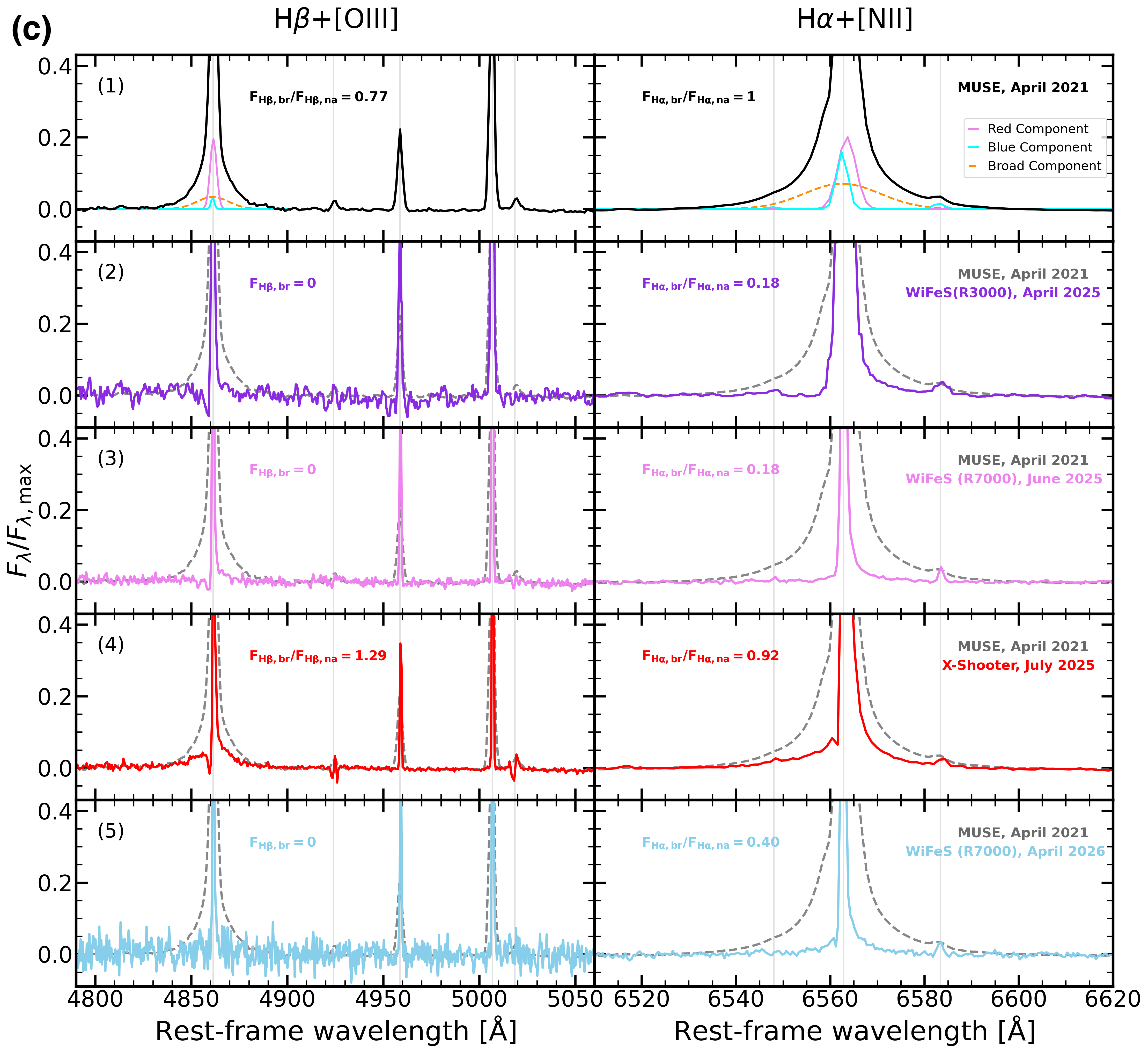}}
\end{center}
\end{figure}
\clearpage
\captionof{figure}{\textbf{Multi-epoch observations of UGCA320-IMBH.} \textbf{a,} Light curves of UGCA320-IMBH collected from Pan-STARRS (diamonds) and ZTF (circles) in the AB magnitude system. Small symbols represent individual exposures and large symbols show one-year median magnitudes; dashed lines indicate the median magnitude in each band. Arrows mark the epochs of the observations presented in this work. The inset shows the \textit{Gaia} light curves in Vega magnitudes. \textbf{b,} SED of UGCA 320-IMBH. Red and green symbols represent observations before the MUSE epoch and near the X-Shooter epoch, respectively. The dashed red curve shows the best-fitting model to the earlier photometry, whereas the dashed green curve shows the fit to the same UV photometry combined with the more recent optical photometry. The model comprises an intrinsic AGN power-law continuum with spectral index $\alpha$, dust attenuation with $A_V=0.1$~mag, and gas attenuation. In the gas attenuation model, $\tau_{\rm b}\gg1$ denotes a large optical depth, $f_{\rm c}$ is the gas covering fraction, and $H(\lambda_{\rm B}-\lambda)$ is a step function that applies attenuation only at $\lambda<\lambda_{\rm B}=3646$\,\AA. \textbf{c,} Multi-epoch spectra of UGCA320-IMBH. The left and right are H$\beta$+[O\,{\sc iii}] and H$\alpha$+[N\,{\sc ii}] spectral windows, respectively, normalized by their peak flux. Panel (1) shows the 2021 MUSE spectrum and its multi-component decomposition, and panels (2)–(5) compare the subsequent WiFeS and X-Shooter spectra (thick colored lines) with the MUSE spectrum (black dashed lines). Faint vertical lines mark the wavelengths of H$\beta$, Fe\,{\sc ii}, the [O {\sc iii}] doublet, H$\alpha$, and the [N {\sc ii}] doublet. The broad-to-narrow Balmer line ratios are listed in each panel.}
\label{fig:change} 

\clearpage

\bibliography{ms}{}

\clearpage

\setcounter{page}{1}
\setcounter{figure}{0}
\setcounter{table}{0}
\renewcommand{\thefigure}{\arabic{figure}}
\renewcommand{\thetable}{\arabic{table}}

\begin{center}
{\bf \Large \uppercase{Methods} }
\end{center}

\section{Multi-Epoch Spectroscopic Observations}\label{sec:obs}
Spectroscopic observations were obtained with the Multi Unit Spectroscopic Explorer (MUSE) \cite{Bacon10} mounted on the Very Large Telescope (VLT) in April 2021, the Wide Field Spectrograph (WiFeS) \cite{Dopita10} on the Australian 2.3 m telescope (ANU-2.3m) in April and June 2025 and April 2026, and VLT/X-shooter \cite{van01} in July 2025. The corresponding pointings are shown in Fig.~\ref{fig:img}b, and a summary of the observations is provided in Extended Data Table~\ref{tab:obs_info}.

\subsection{VLT/MUSE Observations} \label{subsec:muse_obs}
VLT/MUSE observations of UGCA 320 were obtained as part of the Dwarf Galaxy Integral-field Survey \cite{LX26} (DGIS; PI: Yong Shi, co-PI: Fuyan Bian). The observations were carried out in Wide Field Mode without adaptive optics (WFM-NOAO-E) under clear conditions on 22 and 23 April 2021 (program ID: 105.20GY.002). The observations covered the central $1'\times1'$ region of UGCA 320. In this configuration, the spectra cover 4600 to 9350\,\AA, with a spectral resolution of $R\sim$1770 at 4800\,\AA\ and $R\sim$2500 at 6500\,\AA. The final datacube was reconstructed with a spatial sampling of $0.4''\times0.4''$ per spaxel. The point-spread function (PSF) of the final datacube was measured by fitting a Moffat profile to a point source within the field of view, yielding a FWHM of 1.02$''$, corresponding to a physical scale of 28 pc \cite{LX26}. The total on-source integration time was 1.2 h, comprising eight exposures of 560 s each. Off-target sky exposures of 120 s were obtained between two science exposures. One science exposure was excluded during data reduction owing to poor data quality, resulting in a final integration time of 3920 s.

\subsection{ANU-2.3m/WiFeS observations} \label{subsec:wifes_obs}
WiFeS is a dual-beam image-slicing integral-field spectrograph (IFS) that provides a contiguous FoV of $25''\times38''$ covering the IMBH. The final data cubes were reconstructed with a spatial sampling of $1''\times1''$ per spaxel. At the RT560 dichroic setting, the B3000/R3000 configuration provides wavelength coverage from 3200 to 9800 \AA, whereas the B7000/R7000 configuration covers 4180 to 7060\,\AA. The corresponding spectral resolutions are approximately $R=3000$ and $R=7000$, respectively. 

Observations were prepared using the WiFeS Observation Preparation Tool and submitted as individual observing blocks (OBs). Owing to the dynamic scheduling of the facility, the April to June observations were executed through multiple OBs distributed over several nights. Each OB was observed in nod-and-shuffle mode, in which the science target and blank sky were observed with equal exposure times while the CCD charge was shuffled synchronously. Two independent sky positions located away from UGCA 320 were used to construct the final sky spectrum. The on-source time of one OB is 1200 s. The observations were obtained under observing constraints requiring a seeing better than $2''$, a lunar illumination fraction below 40\%, and a minimum angular separation of 30$^{\circ}$ from the Moon.

\subsection{VLT/X-Shooter observations} \label{subsec:xshooter_obs}
Follow-up spectroscopy was obtained with X-shooter mounted on the VLT as part of a Director's Discretionary Time (DDT) program (PI: X.Li; program ID: 115.29FU.001). X-shooter is a medium-resolution spectrograph that simultaneously records spectra in three arms (UVB, VIS, and NIR) with a slit length of 11$''$, providing wavelength coverage from approximately 3000 to 24800\,\AA. The observations were performed using slit widths of 0.8$''$, 0.9$''$, and 0.9$''$ in the UVB, VIS, and NIR arms, providing spectral resolutions of $R\sim$6700, 8900, and 5600, respectively. The pointing position was determined from \textit{Gaia} astrometry and was centered on the unresolved IMBH counterpart. The observations were executed under constraints requiring seeing $<0.9''$, clear sky conditions, an airmass $<1.6$, a fractional lunar illumination $<30\%$, and a minimum target--Moon separation of $30^\circ$. An object–sky–object observing sequence was adopted with ``Fixed offset to sky (SLT)'' mode, in which the sky frames were obtained through separate telescope pointings to a nearby blank region of IMBH, and with equal exposure times assigned to the science and sky frames.

\section{Spectroscopic Data Reduction} \label{sec:data}
\subsection{VLT/MUSE Reduction} \label{subsec:muse_data}
The datacube was produced by the DGIS team \cite{LX26}; a summary of the processing specific to UGCA 320 is provided here. UGCA 320 was exposed 8 times, with each frame equal to a 560-s observation. Raw data were reduced using the standard MUSE pipeline under the \texttt{esoreflex} environment \cite{Freudling13}. Off-target skys were subtracted during the pipeline processing. Subsequent calibrations were performed on the pipeline-reduced individual exposures.  

First, each single-exposure datacube was reprojected onto a common spatial grid using the \texttt{PYTHON} package \texttt{reproject.reproject$\_$interp}, after aligning their World Coordinate System (WCS). By adopting median spectra within the same apertures across all exposures, the exposure at 22 April 00:22 has negative flux levels, so that was excluded from analysis. 

Second, several blank regions in the datacubes were selected as residual background. Their median spectrum was subtracted from the entire datacube as a residual sky, and their standard deviation was quadratically added to the datacube errors. The blank regions were also checked using IRAC 3.6 $\mu$m and Pan-STARRS \cite{Chambers16} broadband images.

Third, $u$, $g$, $r$, and $i$-band  Pan-STARRS images were used to construct a broadband SED within the extraction aperture. For each MUSE exposure, synthetic broadband photometry was generated by convolving the datacube with the corresponding Pan-STARRS filter transmission curves. A reference exposure was selected based on its close agreement with the Pan-STARRS SED shape and its smooth spectral continuum. For each exposure, the median spectrum within the aperture was smoothed using 150\,\AA\ bins to trace the large-scale continuum shape. The ratio between the smoothed spectrum and that of the reference exposure was fitted with a first-order polynomial below 8800\,\AA. The resulting correction was applied to the entire datacube, including both flux and variance, yielding a consistent continuum shape and flux scale across all exposures while preserving narrow spectral features.

Fourth, all corrected datacubes were combined with equal weights. The flux scale of the resulting datacube was calibrated against Pan-STARRS broadband photometry by comparing synthetic and observed broadband SEDs, and the derived first-order correction was applied to the entire datacube, including the variance cube.

Finally, the combined datacube was rebinned to a spatial scale of 0.4$''$ (14 pc), astrometrically aligned to the Pan-STARRS, and corrected for Milky Way (MW) foreground extinction using the Calzetti extinction curve \cite{Cardelli89} with A$_{\rm v}$ = 0.08 (ref-\citenum{Dale09}) and R$_{\rm v}$ = 3.1.

\subsection{ANU2.3-m/WiFeS Reduction} \label{subsec:wifes_data}
The detailed WiFeS data reduction process is the same as that of the DGIS team \cite{LX26}. We briefly summarize it here.

The WiFeS data were reduced using the \texttt{pyWiFeS} pipeline \cite{Childress14}, including cosmic-ray removal, wavelength calibration, telluric correction, atmospheric extinction correction, flux calibration, and sky subtraction through the nod-and-shuffle mode. Barycentric corrections were applied using the \texttt{PYTHON} package \texttt{astropy}, and the wavelength calibration was verified using sky and nebular emission lines. The variance arrays were validated against the observed sky fluctuations and found to be consistent within ($\sim$20\%).

For each grating configuration, the sky-subtracted datacubes from individual exposures were projected onto a common wavelength grid in a flux-conserving manner and combined using exposure-time weighting. The resulting blue- and red-arm datacubes were then resampled onto a common logarithmic wavelength grid corresponding to Nyquist sampling. The flux of the red-arm datacube was normalized to the blue-arm and then merged into a single datacube. 

Finally, the initial WCS was added to the datacubes using the input position and then matched to the SkyMapper $r$-band image \cite{Onken24} with an accuracy of $\lesssim 1''$. The same MW extinction correction was also applied to the final datacube.

\subsection{VLT/X-Shooter Reduction} \label{subsec:xshooter_data}

The X-shooter data were reduced using the X-shooter pipeline under the \texttt{esoreflex} environment. Although the observations were obtained in the SLT mode, the offset sky position contains a point source located $\sim7''$ southwest of the IMBH. To avoid potential contamination from this source, the data were processed using the staring-mode workflow rather than the standard fixed offset sky-subtraction procedure. The pipeline performed bias and dark correction, flat-fielding, wavelength calibration, order merging, flux calibration, and telluric correction, producing two-dimensional spectra for each exposure. A residual background subtraction was also applied by the pipeline. Inspection of source-free regions confirmed that the background level was consistent with zero within the noise. One-dimensional spectra were then extracted from a 1$''$ aperture centered on the peak continuum position of the IMBH in each exposure and combined to obtain the final spectrum. The stacked spectrum was also corrected for the MW extinction.

\section{Spectral Analysis} \label{sec:spec_analy}
To characterize the gas kinematics and emission-line properties associated with the IMBH, we performed spectral fitting on both the spatially resolved MUSE datacube and the integrated multi-epoch spectra. The MUSE data were analyzed on a spaxel-by-spaxel basis to map the spatial distribution of the distinct kinematic components, while the stacked spectra were fitted with a more detailed model to constrain their line profiles and physical properties. The corresponding fitting parameters derived from the multi-epoch spectra are summarized in Extended Data Table~\ref{tab:spec_info}.

\subsection{MUSE Pixel-based Two-Component Spectral Fitting} \label{subsec:pix_analy}
To investigate the spatially resolved gas kinematics (Fig.~\ref{fig:img}), we performed emission-line fitting on a spaxel-by-spaxel basis. The stellar continuum was first modeled and removed using \texttt{pPXF} \cite{Cappellari17}, adopting \texttt{HR-PYPOPSTAR} stellar templates \cite{Milln-Irigoyen21} together with nebular emission and 3rd-order polynomial functions (\texttt{degree=3, mdegree=-1}).

After continuum subtraction, the [S\,{\sc ii}]$\lambda\lambda$6716,6731 doublet was fitted first using Gaussian profiles with \texttt{scipy.optimize.curve\_fit}. The two [S\,{\sc ii}] lines were required to share the same velocity and velocity dispersion and were used to define the kinematic template for the narrow-line emission. 

A two-step emission-line fitting procedure was adopted. In the first step, each emission line was modeled with a single Gaussian component. The H$\beta$+[O\,{\sc iii}]$\lambda\lambda$4959,5007 and H$\alpha$+[N\,{\sc ii}] $\lambda\lambda$6548,6583 complexes were fitted simultaneously. The [O\,{\sc iii}] and [N\,{\sc ii}] doublets were constrained to have fixed flux ratios of 1:3, with their velocities and velocity dispersions tied to the narrow-line kinematic template derived from [S\,{\sc ii}]. The velocities and velocity dispersions of the H$\alpha$ and H$\beta$ narrow components were allowed to vary by up to 25\% relative to the template values. In the second step, an additional broad Gaussian component was introduced for H$\alpha$, H$\beta$, and the [O\,{\sc iii}] doublet. A broad component existed only if its flux was detected with S/N$>$7, its intrinsic velocity dispersion exceeded 300 km s$^{-1}$ after correction for instrumental broadening, and its inclusion reduced the reduced-$\chi^2$ by at least 20\%.

The flux of narrow components was corrected for dust extinction by the narrow line flux ratio of H$\alpha$-to-H$\beta$ and the Small Magellanic Clouds (SMC)-bar extinction curve \cite{Gordon03}, using the intrinsic ratio of (H$\alpha$/H$\beta$)$_{0}$=2.86 and R$_v$ = 2.74.

\subsection{Multi-component Fitting of the Stacked Spectra} \label{subsec:stack_analy}
For the stacked spectra, the continuum was modeled with a single power-law component. Stellar templates were not included because the previous \texttt{pPXF} analysis indicated a negligible stellar contribution within the extraction apertures. In addition, the Balmer absorption features are shown below to be non-stellar in origin and were therefore modeled explicitly as separate absorption components rather than through stellar-continuum fitting.

For MUSE observation, the integrated spectra of IMBH were produced by stacking the spaxels with a detected broad H$\alpha$ component according to previous pixel-based fittings, and the stacked spectra were used to derive the BH mass and corresponding properties. For WiFeS, spectra within an aperture corresponding to one PSF ($\sim 2''$) were combined. For X-shooter, the spectrum was extracted from a $1''$ aperture centered on the continuum peak of the IMBH as we described before. We verified that the derived spectral properties were insensitive to modest changes in the extraction aperture, with consistent results obtained when the aperture radius was varied within $\sim3''$.

For the MUSE spectrum, no Balmer absorption component was detected. The [S\,{\sc ii}] and Fe\,{\sc ii} emission lines were first fitted independently with single Gaussian profiles. The [S\,{\sc ii}] emission exhibits a lower velocity and was adopted as the template for the blueshifted narrow component, whereas the Fe\,{\sc ii} emission defines the redshifted component. The H$\beta$+[O\,{\sc iii}] and H$\alpha$+[N\,{\sc ii}] complexes were then fitted simultaneously. The [O\,{\sc iii}] and [N\,{\sc ii}] doublets were tied to the blueshifted [S\,{\sc ii}] kinematics, while the redshifted narrow component was introduced with free Gaussian parameters. A broad Balmer component was subsequently added. Owing to the large extraction aperture, the velocity gradient is spatially blended, and no distinct second component is required for the [O\,{\sc iii}] doublet.

For the WiFeS spectra, the S/N was insufficient to reliably constrain the Balmer absorption component and the Fe\,{\sc ii} emission. Therefore, firstly, following the MUSE analysis, the narrow Balmer component was tied to the kinematics of the [S\,{\sc ii}] component. A second narrow component and subsequently a broad Balmer component were introduced sequentially, with their kinematic and flux parameters left free. Additional components were retained only if their fluxes were detected with S/N$>$7 and their inclusion reduced the reduced-$\chi^2$ by at least 20\%. Under these criteria, the H$\beta$ profile was adequately described by a single narrow component, whereas additional components were required for H$\alpha$.

The X-shooter spectra were modeled using the same kinematic framework as the MUSE spectra. The blueshifted narrow component was tied to the kinematics of the [S\,{\sc ii}] emission, while the redshifted component was constrained by the Fe\,{\sc ii} emission lines. However, strong Balmer absorption features are present and must be accounted for during the fitting. To avoid biasing the emission-line measurements, the absorption regions were initially masked while fitting the emission components. The absorption profiles were then modeled from the residual spectra obtained after subtracting the best-fitting emission-line model.

\section{Astrometric Alignment between MUSE and HST}

Detailed astrometric calibration was performed to align the MUSE broad-H$\alpha$ flux map with the \textit{HST} ACS/WFC3 F606W and F814W images using the \texttt{PYTHON} package \texttt{Astroalign} \cite{Beroiz20}, adopting Pan-STARRS broad-band images as the absolute astrometric reference frame.

First, the \textit{HST} images were convolved to a common spatial resolution of 1$''$ to match the seeing of the Pan-STARRS data. The PSF-matched HST images were then registered to the Pan-STARRS $r$- and $i$-band images using \texttt{Astroalign}, which identifies common bright sources in both datasets and derives a geometric transformation matrix to align the source positions. Second, a pseudo $r$-band image was constructed from the MUSE datacube by convolving the spectra with the Pan-STARRS $r$-band filter curve. This MUSE $r$-band image was then registered to the Pan-STARRS reference frame using the same procedure, and the derived transformation was subsequently applied to the MUSE broad-H$\alpha$ flux map.

This procedure ensures that both the \textit{HST} and MUSE datasets are placed on a consistent Pan-STARRS-based astrometric frame, resulting in a relative alignment accuracy of $\sim0.05''$ between MUSE and HST. Finally, this accurate registration allows a direct spatial comparison between the MUSE broad-H$\alpha$ emission and the \textit{HST} point-source morphology, confirming that the peak of the broad-H$\alpha$ component is spatially consistent with a compact source in the HST images within the astrometric uncertainty.

\section{Light Curves}
The $g$-, $r$-, and $i$-band light curves before 2016 were collected from the Panoramic Survey Telescope and Rapid Response System (Pan-STARRS) DR2 Detection catalog using the \texttt{psfFlux} measurements. After 2017, photometric measurements were obtained from the Zwicky Transient Facility (ZTF) DR23 catalog \cite{Masci19} through the NASA/IPAC Infrared Science Archive (IRSA) \cite{IRSA22}. No monitoring observations were available from either Pan-STARRS or ZTF between 2015 and 2017. To bridge this gap, \textit{Gaia} DR3 Epoch Photometry (https://doi.org/10.17876/gaia/dr.3/55; source ID 3511748501697415168) was collected in the G, BP, and RP bands. The \textit{Gaia} photometry was used in the Vega magnitude system. The median magnitude within each year was adopted as the representative value. The associated uncertainties were estimated by combining the individual photometric uncertainties and the standard deviation of all measurements obtained within the corresponding year.

\section{Collections of X-ray-to-Optical Photometry}
The X-ray-to-optical SED of IMBH was collected from: X-ray from \textit{Chandra} \cite{Garmire03}, far and near ultraviolet (UV) from \textit{GALEX} (\textit{Galaxy Evolution Explorer}) \cite{Martin05}, near-UV from \textit{Swift} \cite{Roming05}, $u$ band photometry from CFHT, and optical band from 4000\AA\ to 10,000\AA\ from  Pan-STARRS, SkyMapper \cite{Onken19}, and \textit{HST}. Photometry was corrected for the MW foreground extinction using the \texttt{PYTHON} package \texttt{extinction}. The Cardelli extinction curve \cite{Cardelli89} are adopted with $R_v$ = 3.1 and dust attenuation $A_v$ = 0.24 \cite{Dale09}.

{$\bullet$ \it Chandra}: A \textit{Chandra} observation of the IMBH field was obtained in December 2024 under program 26700179 (PI: J.~Wang) using the Advanced CCD Imaging Spectrometer spectroscopic array (ACIS-S). The observations cover the 2--10 keV energy range with a total exposure time of 50 ks, distributed among five individual exposures. The observations sensitivity achieved a flux upper limit of $5\times10^{-15}\,{\rm erg\,s^{-1}\,cm^{-2}}$ (Extended Figure data~\ref{fig_Lx}).

{$\bullet$ CFHT:} CFHT $u$-band observations were carried out in May 2025 under program 25AS02 (PI: X.~Li) using the MegaCam instrument. The observations were executed in two dithering patterns to improve background sampling and mitigate detector artifacts: a \texttt{DP8} pattern with 1$''$ step size and a second \texttt{DP8} pattern with 1.5$''$ step size. Each dithering pattern was executed as a separate OB, resulting in a total of two OBs and 16 individual exposures. The raw data were processed using the \texttt{Elixir} pipeline \ cite {Magnier04}, which performs bias subtraction, flat-field correction, astrometric calibration, and initial photometric zero-point determination. For each individual frame, aperture photometry of the IMBH was performed, and the final $u$-band flux was obtained by stacking the measurements using exposure-time weighting. A circular aperture of 2$''$ radius was adopted, with the background estimated from an annulus defined in the outer region around the source and subtracted accordingly.

{$\bullet$ \it GALEX}: The photometry was derived from archival imaging data obtained through the Gest Investigators (GI) dataset. The processed UV images were retrieved from the \textit{Spitzer} Local Volume Legacy survey \cite{Dale09} (https://irsa.ipac.caltech.edu/data/SPITZER/LVL/). The IMBH is located in a relatively isolated region, allowing it to be spatially resolved from the host galaxy emission in \textit{GALEX} images, with a PSF of approximately 5$''$ in the far-UV and 2.5$''$ in the near-UV.

Aperture corrections were derived using a set of bright, isolated point sources in the field. For each source, the centroid was defined by the peak pixel, and aperture photometry was performed using concentric circular apertures with radii ranging from 1.5$''$ to 9$''$. Background subtraction was applied using local background maps extracted at the same positions. The aperture correction was defined as the magnitude difference between the largest aperture and the total flux, $\Delta m = m_{r<9''} - m_{r=9''}$. The resulting correction as a function of radius was normalized to the values provided in the \textit{GALEX} technical documentation. The final aperture correction curve was taken as the median of all selected point sources, and its uncertainty was estimated by combining the standard deviation among sources and their individual photometric uncertainties.

Photometry of the IMBH counterpart was performed using a fixed aperture centered on the peak of the broad H$\alpha$ emission. A radius of 3.8$''$ was adopted, and aperture corrections were applied accordingly ($\Delta m_{\mathrm{FUV}} = 1.04 \pm 0.11$ mag and $\Delta m_{\mathrm{NUV}} = 0.41 \pm 0.02$ mag). Background subtraction was performed using matched apertures on the corresponding background maps. The total photometric uncertainty includes contributions from photon noise, background fluctuations, zero-point calibration uncertainties (0.01 mag), and centroiding errors. Photometric errors due to background and photon noise were estimated following \cite{Gil05, Gil07}. Centroiding uncertainties were assessed using a Monte Carlo approach in which both the aperture radius ($1.5''\sim3.8''$) and centroid position (within one MUSE PSF of 1.08$''$) were varied, and the photometry was repeated 1000 times.

{$\bullet$ \it Swift:} The near-UV photometries were also obtained from the \textit{Neil Gehrels Swift Observatory} Ultraviolet/Optical Telescope (UVOT) in the UVW2, UVM2, and  UVW1 bands. The processed images were retrieved from the \textit{Swift} Data Archive hosted by the High Energy Astrophysics Science Archive Research Center (HEASARC). The UVOT images have a typical spatial resolution of $\sim$2.5--3.0$''$. Photometry was performed using a fixed aperture centered on the position of the IMBH identified from the broad H$\alpha$ emission. Source fluxes were extracted within a circular aperture, while the local background was estimated from a surrounding annulus centered on the same position and subsequently subtracted.

{$\bullet$ Pan-STARRS:} Point-source photometry was taken from the archival PS1 DR2 \texttt{MeanObject} catalog, using the \texttt{MeanPSFMag} measurements.

{$\bullet$ SkyMapper:} Photometry was obtained from the SkyMapper DR4 catalog, adopting the PSF-based magnitudes provided in the \texttt{$\ast\_{\rm psf}$} columns, where ``\texttt{$\ast\_$}" denotes the corresponding photometric band.

{$\bullet$ \it HST:} The \textit{HST} ACS/WFC3 F606W and F814W images were obtained from Mikulski Archive for Space Telescopes (MAST). To enable consistent comparison with ground-based observations, the images were PSF-matched to the Pan-STARRS imaging by convolution with the corresponding Pan-STARRS point-spread function. Background emission was estimated and removed using the \texttt{PYTHON} package \texttt{photutils} with its \texttt{background.MMMBackground} algorithm.

Aperture corrections ($\Delta m$) for the PSF-matched images were derived using a set of bright, isolated point sources in the field. For each source, aperture photometry was performed using a range of circular apertures with radii from 0.3$''$ to 3$''$. The resulting growth curves were used to determine the aperture corrections as a function of radius. Photometry of the IMBH counterpart was performed using a fixed aperture with a radius of 1.3$''$, and aperture corrections were applied accordingly ($\Delta m_{\mathrm{F606W}} = 0.06 \pm 0.03$ mag and $\Delta m_{\mathrm{F814W}} = 0.04 \pm 0.02$ mag). The total photometric uncertainties include contributions from photon noise, background fluctuations, and uncertainties in the aperture correction.

\section{SED fittings}
\subsection{AGN+Dense Gas}
The SED was modeled using an AGN power-law continuum modified by Balmer-continuum absorption and internal dust attenuation. The intrinsic continuum was represented by a power law, $F_{\lambda}\propto\lambda^{\alpha}$, where $\alpha$ denotes the spectral index. Motivated by the interpretation of Refs-\citenum{Inayoshi25, Ji25}, in which the blueshifted Balmer absorption features observed in little red dots were attributed to partially covering optically thick gas, the continuum blueward of the Balmer edge ($\lambda_{\rm B}=3646$\,\AA) was modeled using a partial-covering attenuation model \cite{Juodzbalis24}: $T_{\rm B}(\lambda)=1-f_{\rm c}(1-e^{-\tau_{\rm b}})H(\lambda_{\rm B}-\lambda)$, where $f_{\rm c}$ is the covering fraction, $\tau_{\rm b}$ is the effective optical depth. The Heaviside step function $H(\lambda_{\rm B}-\lambda)$ restricts the attenuation to wavelengths shorter than the Balmer edge. In the optically thick limit ($\tau_{\rm b}\gg1$),  the component transmitted through the absorber becomes negligible, such that the transmission factor blueward of the Balmer edge reduces to $1-f_{\rm c}$.

Internal dust attenuation was incorporated using the SMC-bar attenuation curve with a fixed attenuation of $A_{\rm V}=0.1$ mag. Under the optically thick approximation, the model can be written as $F_{\lambda,{\rm mod}}=A\lambda^{\alpha} \left[1-f_{\rm c}H(\lambda_{\rm B}-\lambda)\right] 10^{-0.4A_{\lambda}}$, where $A$ is the normalization constant and $A_{\lambda}$ was calculated from the SMC-bar attenuation curve normalized to $A_{\rm V}=0.1$ mag. The free parameters were $A$, $\alpha$, and $f_{\rm c}$. The fitting was performed in logarithmic flux space using the \texttt{PYTHON} package \texttt{scipy.optimize.curve\_fit}. 

The model was first fitted to photometries obtained before the MUSE observations, including data from \textit{GALEX}, \textit{Swift}, Pan-STARRS, SkyMapper, and \textit{HST}.  The fitting was subsequently repeated using more recent observations, including CFHT $u$-band observations and the mean flux measured during the final year of ZTF monitoring. As no updated UV observations are available, previous UV photometry was retained to constrain the ultraviolet portion of the SED, although the UV flux may be lower than the adopted values owing to the observed long-term fading of the source.

\subsection{Stellar Models}
For the stellar SED fitting, the ATLAS9 stellar templates \cite{Castelli03} were used. The ATLAS9 stellar templates cover various types of stellar atmospheres, from cold dwarf stars to hot evolving O/B stars. The model grid spans effective temperatures ($T_{\rm eff}$) from 3,500 k to 50,000 k, logarithmic of surface gravities ($\log g$) from 0.0 to 5.0 cm\,s$^{-2}$, and stellar metallicities ($[M/H]$) from $-2.5$ to $+0.5\,Z_{\odot}$. 

Based on the ATLAS9 stellar templates, the SED fitting was performed using a linear combination of two single stellar population (SSP) templates. The model spectra were then convolved with the transmission curves of all corresponding photometric filters used in observations to produce synthetic broadband fluxes. These were converted to flux densities and normalized by their median values to focus the fitting on the SED shape rather than the absolute normalization.

Dust attenuation was applied over a grid of visual extinctions $A_V = 0-1.0$ mag. The fitting was performed in logarithmic flux space, and the likelihood was evaluated using a reduced-$\chi^2$ statistic that accounted for photometric uncertainties.

For each pair of SSP templates, the linear coefficients of the two components were optimized using bounded least-squares fitting, with the constraint that both coefficients were non-negative. The best-fitting solution was determined by minimizing $\chi^2$ over the full grid of template pairs and extinction values, yielding the preferred combination of stellar population parameters, dust attenuation, and relative component contributions.

\section{Color-Magnitude Diagram and Isochrone Fittings}
We note that ATLAS9 stellar atmosphere models do not directly provide bolometric luminosities, which need to be implemented with stellar evolutionary tracks. Therefore, the physical properties of the composite stellar components inferred from the SED fitting were further constrained through isochrone fitting in the color–magnitude diagram (CMD). Stellar isochrones were generated using the MIST (Modules for Experiments in Stellar Astrophysics Isochrones and Stellar Tracks) evolutionary library \cite{Choi16}, implemented via the \texttt{PYTHON} package \texttt{ArtPop} \cite{Greco22}. MIST library covers metallicity of [Fe/H]/[Fe/H]$_{\odot}$ = $-0.5$, $-0.25$, and 0, and stellar ages from 10 Myr to 1 Gyr. 

We first constructed the observed CMD from HST F606W and F814W photometry. The position of the IMBH in the CMD was then matched to MIST isochrones to infer the $T_{\rm eff}$ and bolometric luminosity $L_{\rm bol}$. Similarly, synthetic F606W and F814W magnitudes predicted from the SED-fitting results were independently fitted in CMD space, getting the $T_{\rm eff}$ and $L_{\rm bol}$ of each stellar component. For consistency, we verified that $T_{\rm eff}$ inferred from MIST-based isochrone fitting is consistent with those obtained using ATLAS9-based SED modeling. Finally, the derived stellar parameters were placed on a Hertzsprung–Russell diagram based on ref.~\citenum{Humphreys14} for comparison with bright stars found in the MW.

\section{Column Density of Absorber}

A grid of photoionization models using the \texttt{CLOUDY} code \cite{Ferland2017} was generated to constrain the density of the absorber, assuming an incident power-law continuum and an ionization parameter of $\log U = -1.5$. The hydrogen density $n_{\rm H}$ is varied logarithmically from $10^{5}$ to $10^{12}~\mathrm{cm}^{-3}$ in steps of 0.5~dex, and the total hydrogen column density $N_{\rm H}$ is varied from $10^{22}$ to $10^{25}~\mathrm{cm}^{-2}$, also in steps of 0.5~dex. 
The elemental abundances were set to be 0.05, 0.5, and 1.0$\times Z_\odot$.

Based on \texttt{CLOUDY}, the transmitted continuum and line spectrum emerging from the absorbing gas were calculated. The Balmer-break strength and Balmer absorption-line equivalent widths were then measured directly from the model spectra. The Balmer-break strength was defined as $D_{\rm B}=\frac{F_{\lambda}(\lambda3800\text{--}4000)}{F_{\lambda}(\lambda3500\text{--}3600)}$. The equivalent width of H$\gamma$ absorption was measured from the continuum-normalized spectrum. By comparing the observed values of $D_{\rm B}$ and EW(H$\gamma$) derived from the X-Shooter spectrum with the model predictions, constraints were placed on the $n_{\rm H}\sim10^{11\sim12}\,{\rm cm^{-3}}$ and $N_{\rm H}\sim5\times10^{22}\,{\rm cm^{-2}}$ of the absorbing gas (Extended Data Fig.~\ref{fig_cloudy}), assuming complete coverage of the continuum source ($f_{\rm c}=1$).  However, both the observed SED and the presence of Balmer emission lines indicate that the absorber only partially covers the emitting region. For a given $D_{\rm B}$ and EW(H$\gamma$), introducing a covering fraction requires stronger intrinsic absorption and therefore implies larger values of both $n_{\rm H}$ and $N_{\rm H}$ (Extended Data Fig.~\ref{fig_cloudy}). Consequently, the constraints derived from the $f_{\rm c}=1$ models should be regarded as conservative lower limits. Noted that we did not consider the contribution of internal dust extinction, and the H$\gamma$ absorption feature can be affected by the contamination of emission components. 
 
\clearpage

\clearpage
\setcounter{page}{1}
\setcounter{figure}{0}
\setcounter{table}{0}
\renewcommand{\thefigure}{\arabic{figure}}
\renewcommand{\thetable}{\arabic{table}}
\renewcommand{\figurename}{Extended Data Figure}
\renewcommand{\tablename}{Extended Data Table}

\begin{center}
{\bf \Large \uppercase{Extended Data} }
\end{center}

\begin{table}[H]
\scriptsize
\centering
    \caption{\label{tab:obs_info} Observation Information of UGCA320-IMBH}
\begin{tabular}{lccccccccc}
\hline
\hline
TEL/INS &    MODE   & Program ID &   Date   &  ExpTime/frame$^{\dagger}$ & ObsSeq/OB$^{\ddagger}$   & N\_OB  & Condition & Seeing   & Airmass  \\
    &            &            & yy-mm-dd &   s     &          &         &          &   $"$    &          \\
\hline
VLT/MUSE & WFM-NOAO-E & 105.20GY.002 & 2021-04-22 & 560 (O), 120 (S) &  O-S-O & 2  & Clear & 1.45(0.09) & 0.33(0.05) \\
        &             &              & 2021-04-23 & 560 (O), 120 (S) &        & 2  & Clear & 1.49(0.1)  & 0.41(0.04)  \\
\hline
VLT/X-Shooter$^{\ast}$ & UVB(slit=0.8$"$) & 115.29FU.001 & 2025-07-22 & 424(O,S)&  O-S-O & 2  & Clear & $<$0.9    & 1.113  \\ 
              &                  &              & 2025-07-23 &         &        & 2  & Clear & $<$0.9    & 1.113   \\ 
              & VIS(slit=0.9$"$) &              & 2025-07-22 & 300(O,S)&  O-S-O & 2  & Clear & $<$0.9    & 1.113 \\ 
              &                  &              & 2025-07-23 &         &        & 2  & Clear & $<$0.9    & 1.113 \\
              & NIR(slit=0.9$"$) &              & 2025-07-22 & 480(O,S)&  O-S-O & 2  & Clear & $<$0.9    & 1.113 \\ 
              &                  &              & 2025-07-23 &         &        & 2  & Clear & $<$0.9    & 1.113\\             
\hline
ANU-2.3m/WiFeS$^{\star}$ & RT560+BR3000    &  2513051     & 2025-04-04  & 150(O,S)& s-O-(S-O)$\times$7-s & 2 & ---   & $<$2  & 1.04  \\
               &                 &              & 2025-04-05  &         &                    & 2 &       &       & 1.04  \\
               &                 &              & 2025-04-21  &         &                    & 1 &       &       & 1.06  \\
               &                 &              & 2025-04-24  &         &                    & 2 &       &       & 1.04  \\
               & RT560+BR7000    &  2513051     & 2025-05-23  & 150(O,S)& s-O-(S-O)$\times$7-s & 2 & ---   & $<$2  & 1.07(0.03)  \\
               &                 &              & 2025-05-24  &         &                    & 2 &       &       & 1.03  \\
               &                 &              & 2025-06-15  &         &                    & 1 &       &       & 1.03  \\
               &                 &              & 2025-06-18  &         &                    & 1 &       &       & 1.04  \\
               &                 &              & 2025-06-26  &         &                    & 1 &       &       & 1.04  \\
               &                 &              & 2025-06-27  &         &                    & 1 &       &       & 1.03  \\
               & RT560+BR7000    & 2613031      & 2026-04-24  & 150(O,S)& s-O-(S-O)$\times$7-s & 4 & ---   & $<$2  & 1.04(0.04)  \\
\hline
CFHT/MEGACam      & u-band       & 25AS02  &  2025-05-25   & 95 &  DP8/1$"$+DP8/1.5$"$ & 1 & Thin Cirrus  & 0.82(0.05) & ---\\
\hline
\hline
\end{tabular}
\begin{minipage}{0.95\linewidth}
\scriptsize
$^{\dagger}$ Exposure time per frame. ``O'' denotes target exposure and ``S'' denotes sky exposure.\\
$^{\ddagger}$ Observing sequence within a single observing block (OB).\\
$\ast$ X-Shooter observations lack a seeing monitor, but the seeing is required to be better than 0.9$''$.\\
$\star$ Weather conditions during the WiFeS observations were not recorded; a seeing constraint of  $<2''$ was required.\\
Note: Blank entries indicate that the configuration was unchanged from the preceding row. A dash ``---'' means that no measurement was available.
\end{minipage}
\end{table}

\begin{table}[H]
\scriptsize
\centering
\caption{\label{tab:spec_info} Multi-Epoch Spectral Properties of  UGCA320-IMBH}    
\begin{tabular}{cccccccc}
\toprule
       &  &  \multicolumn{2}{c}{MUSE}  &   \multicolumn{2}{c}{WiFeS}  &   \multicolumn{2}{c}{X-Shooter}  \\
\hline
       &  &  H$\beta$ & H$\alpha$ &  H$\beta$ & H$\alpha$ &  H$\beta$ & H$\alpha$ \\
\cmidrule(lr){3-4} \cmidrule(lr){5-6} \cmidrule(lr){7-8}
$F_{\rm na,b}$ & 1E-16\,$\mathrm{erg/s/cm^2}$ & 5.69$\pm$0.52 & 71.64$\pm$0.16  & 18.22$\pm$0.31  &  27.54$\pm$4.98   &  3.40$\pm$0.11 & 20.10$\pm$0.90  \\
$F_{\rm na,r}$ & 1E-16\,$\mathrm{erg/s/cm^2}$ & 66.26$\pm$0.60 & 141.95$\pm$0.97  & --- &  60.78$\pm$4.26 & 2.10$\pm$0.10  & 10.82$\pm$0.91  \\
$F_{\rm br}$ & 1E-16\,$\mathrm{erg/s/cm^2}$ & 55.95$\pm$1.98 & 214.19$\pm$0.64  &  ---  & 14.93$\pm$1.64  & 7.10$\pm$0.28  &  28.53$\pm$1.46 \\
$v_{\rm br}$ & km/s & -6.72$\pm$1.92 & -11.88$\pm$0.09  &  ---  & 251.68$\pm$29.32  &  30.81$\pm$16.96  &   54.24$\pm$16.95\\
FWHM$_{\rm br}$ & km/s & 1192.73$\pm$52.63 & 857.65$\pm$6.09  &  ---  & 537.35$\pm$37.13  & 947.22$\pm$36.85  & 785.30$\pm$44.82  \\
$v_{\rm na,b}$ & km/s &  \multicolumn{2}{c}{-16.21$\pm$0.02}  & ---  & -8.15$\pm$0.85 & \multicolumn{2}{c}{19.02$\pm$0.84} \\
$v_{\rm na,r}$ & km/s &  12.02$\pm$0.42  & 34.54$\pm$0.16  & --- & 21.54$\pm$0.84  & \multicolumn{2}{c}{76.84$\pm$5.95} \\
FWHM$_{\rm na, r}$ & km/s & 170.13$\pm$1.77 & 162.46$\pm$0.66  & --- & 155.19$\pm$7.26  & \multicolumn{2}{c}{67.31$\pm$14.10}  \\
$\mathrm{[S\,II]}\,6716/6731$  &     & \multicolumn{2}{c}{1.42$\pm$0.03} & \multicolumn{2}{c}{1.36$\pm$0.05 } & \multicolumn{2}{c}{1.38$\pm$0.07 }\\
\bottomrule
\end{tabular}
\begin{minipage}{0.9\linewidth}
\scriptsize
(i) The subscripts ``na'' and ``br'' denote the narrow- and broad-line components, respectively.\\
(ii) The subscripts ``b'' and ``r'' denote the blueshifted and redshifted components, respectively.\\
(iii) The velocity dispersions have been corrected for instrumental broadening.
\end{minipage}
\end{table}

\begin{figure}[H]
\begin{center}
\includegraphics[width=0.4\textwidth]{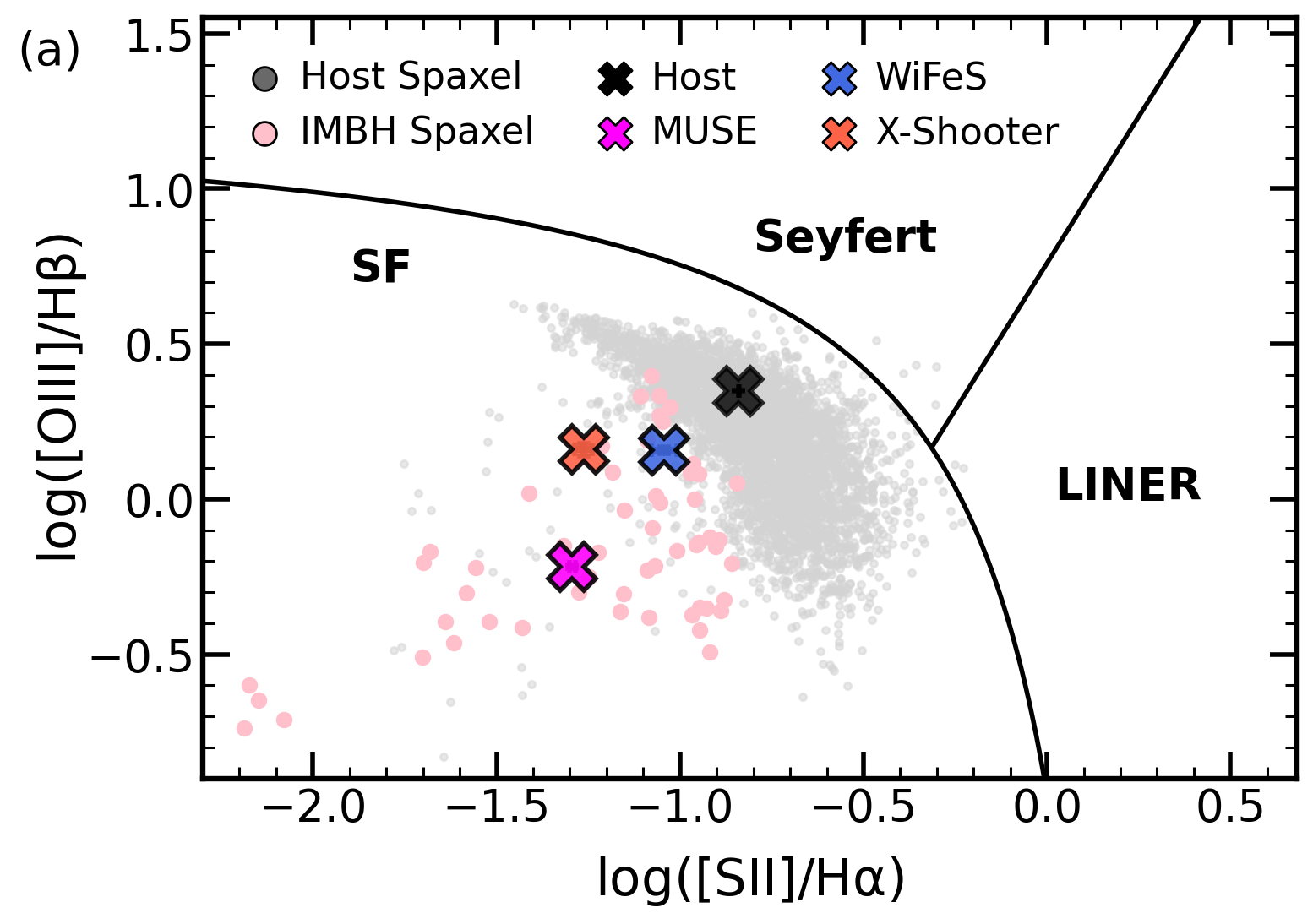} 
\includegraphics[width=0.4\textwidth]{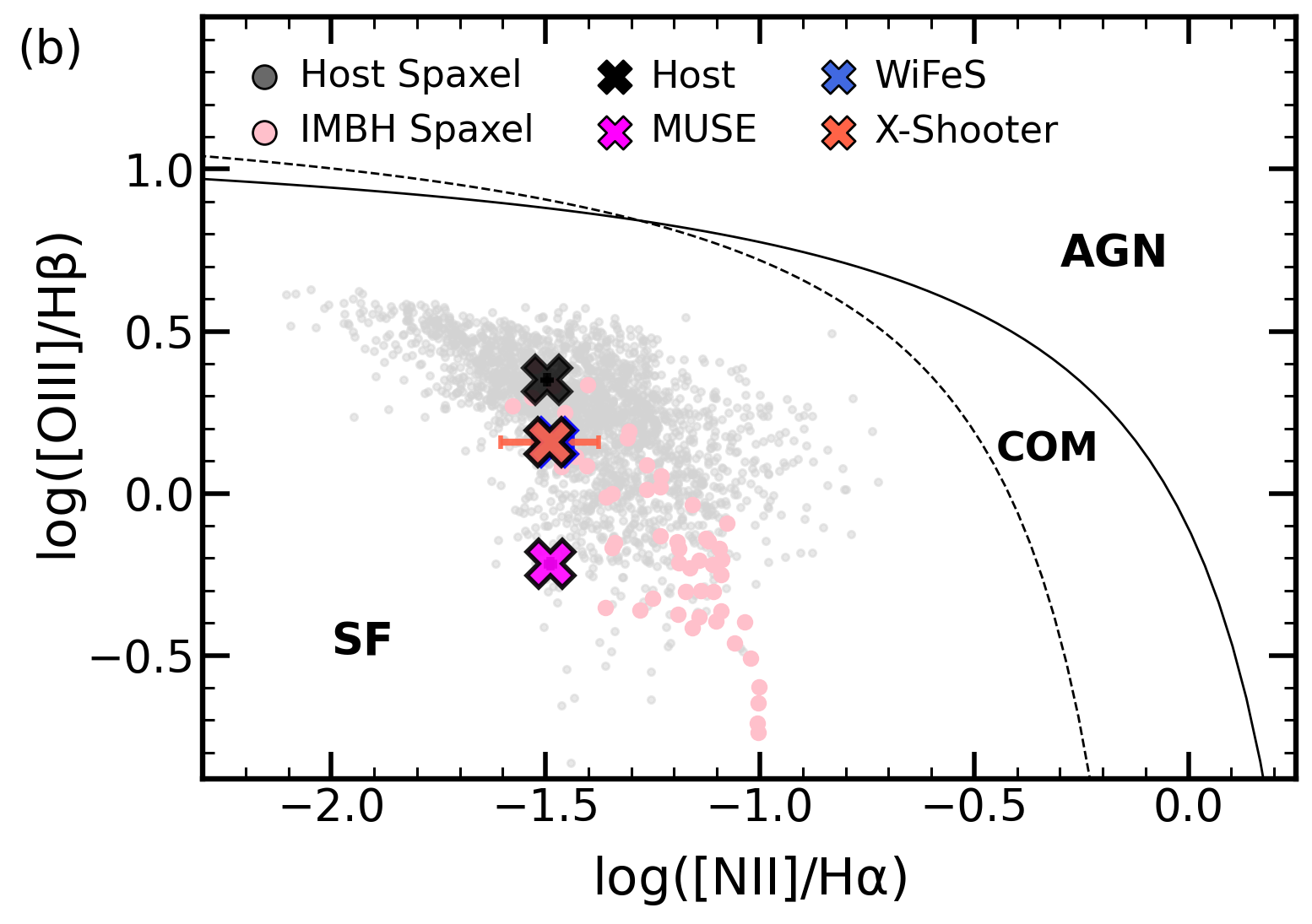} 
\end{center}
\vspace{-4mm}
\caption{\label{fig_bpt}: \textbf{Baldwin–Phillips–Terlevich (BPT) diagnostic diagrams of UGCA320-IMBH and its host galaxy.} {\bf a,} [S {\sc ii}]-BPT diagram. {\bf b,} [N {\sc ii}]-BPT diagram. Grey and pink circles show line ratios measured from individual MUSE spaxels in the host galaxy and the IMBH environment, respectively. Crosses indicate integrated measurements of the host galaxy (black) and the UGCA320-IMBH (magenta) from the MUSE, WiFeS (blue), and X-Shooter (tomato). Solid and dashed curves divide star-forming (SF), composite (COM), Seyfert, and LINER/AGN excitation regimes, which are adopted from ref-\citenum{Kewley01, Kauffmann03, Kewley06}.  }
\vspace{-4mm}
\end{figure}

\begin{figure}[H]
\begin{center}
 \includegraphics[width=0.54\textwidth]{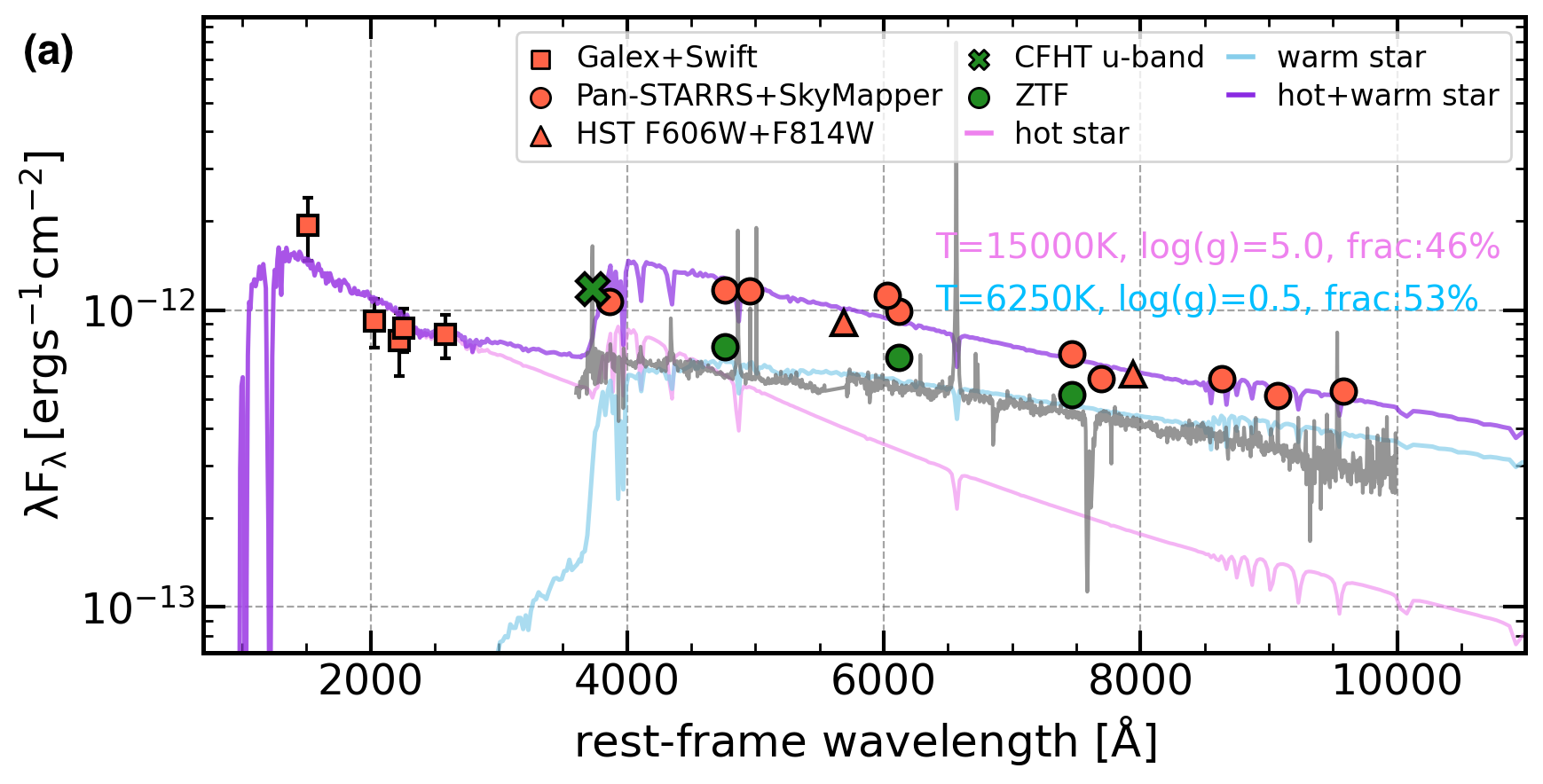} 
\includegraphics[width=0.4\textwidth]{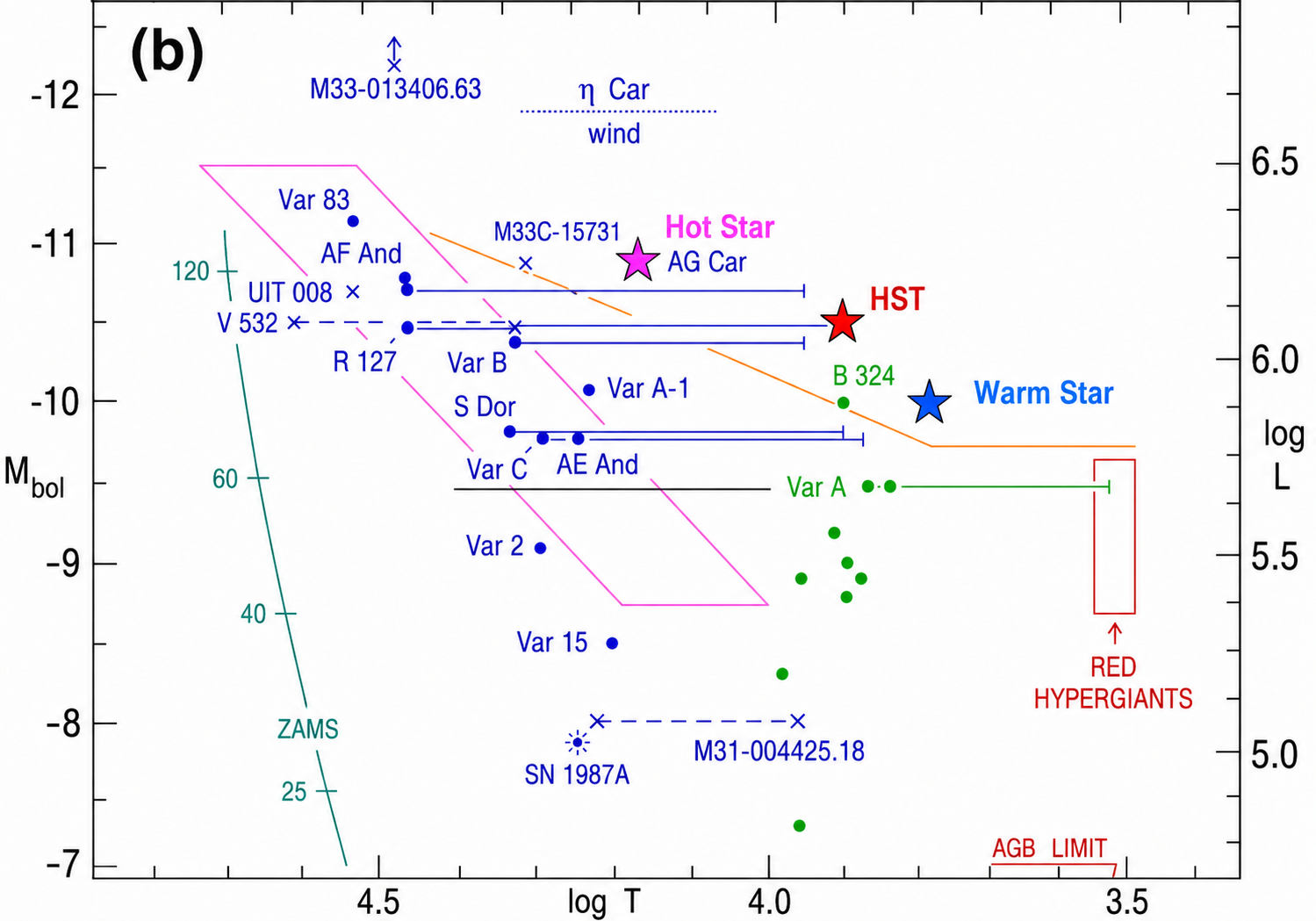}    
\end{center}
\vspace{-4mm}
\caption{\label{fig_star} \textbf{The stellar interpretation of UGCA320-IMBH.} {\bf a,} UV-to-optical SED of UGCA 320-IMBH, corrected for Galactic foreground extinction. Red and green symbols represent observations before the MUSE and near the X-Shooter epochs, respectively. The smoothed X-Shooter spectrum is shown in grey. The purple curve shows the best-fitting stellar model, consisting of a hot-star component (pink) and a warm-star component (blue). {\bf b,} Hertzsprung–Russell diagram adapted from ref.~\cite{Humphreys14}. The red star marks the best-fitting stellar model derived from the HST photometry, while the pink and blue stars indicate the hot and warm stellar templates based on SED fitting. The orange curve shows the empirical upper luminosity boundary of massive stars. }
\vspace{-4mm}
\end{figure}

\begin{figure}[H]
\begin{center}
 \includegraphics[width=0.54\textwidth]{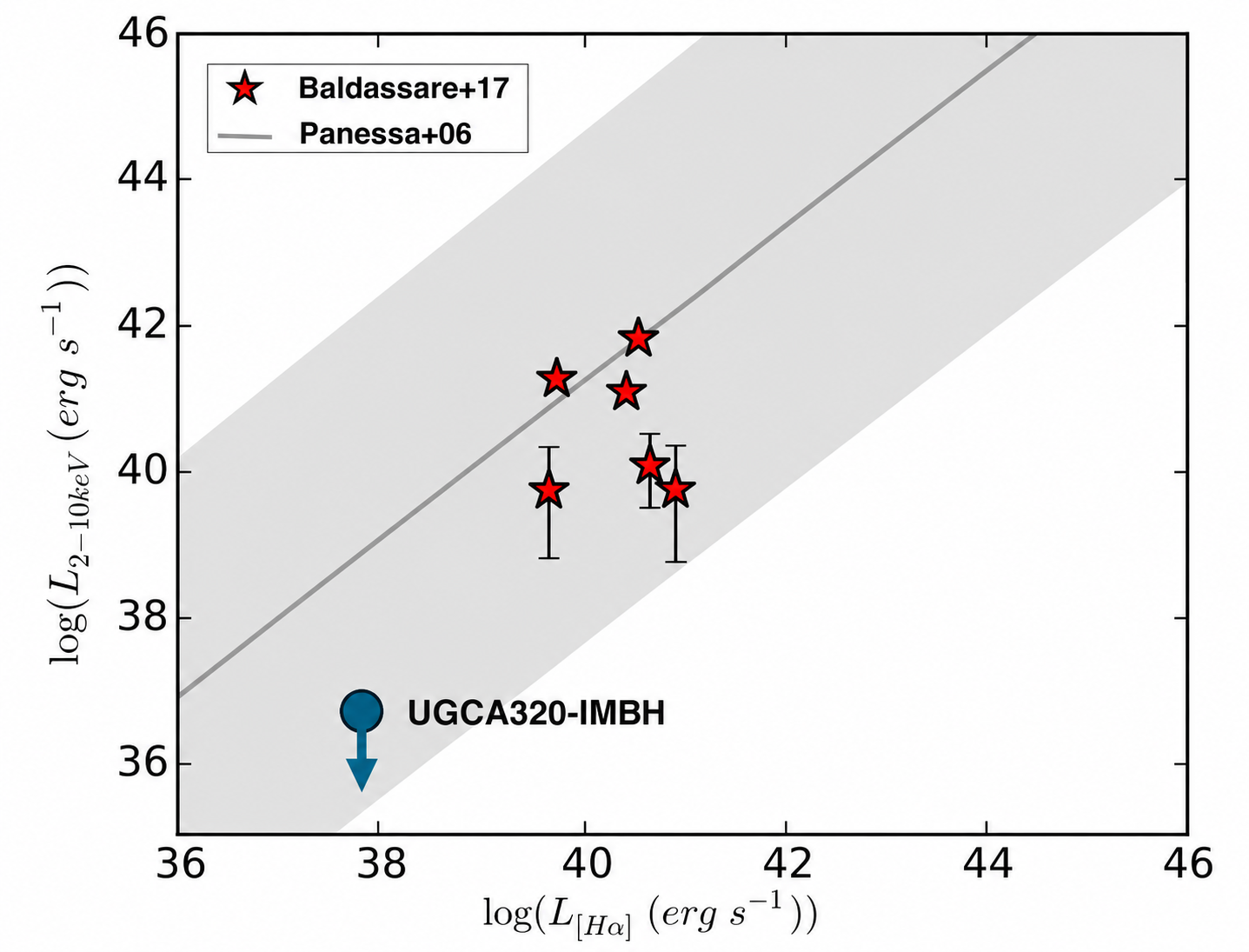}   
\end{center}
\vspace{-4mm}
\caption{\label{fig_Lx} \textbf{X-ray and H$\alpha$ luminosities of broad-line AGN in dwarf galaxies.} The images were adopted from ref-\citenum{Baldassare17}, where red stars with error bars are massive BHs in their sample, and the grey solid line with shaded regions represents the empirical relation of ref-\citenum{Panessa06}. The blue symbol represents the x-ray upper limit of UGCA320-IMBH.}
\vspace{-4mm}
\end{figure}

\begin{figure}[H]
\begin{center}
 \includegraphics[width=0.5\textwidth]{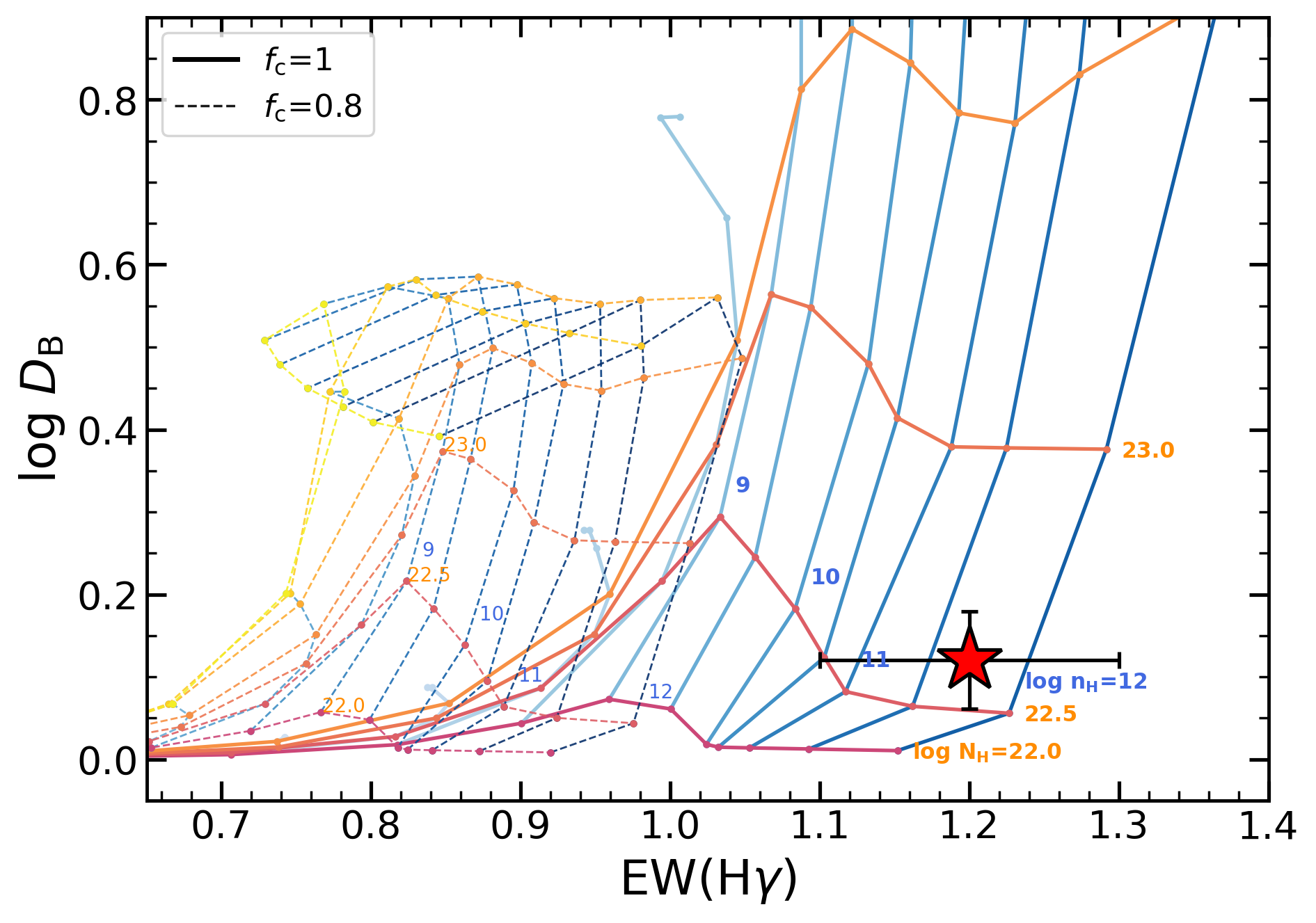}   
\end{center}
\vspace{-4mm}
\caption{\label{fig_cloudy} \textbf{Absorber properties constrained using \texttt{CLOUDY}}. The model predictions are compared with the Balmer break strength ($D_{\rm B}$) in logarithm and H$\gamma$ absorption equivalent width (EW(H$\gamma$)). The Balmer-break strength is defined as $D_{\rm B}=F_{\lambda}(\lambda3800\sim4000)/F_{\lambda}(\lambda3500\sim3600)$. The model grid spans hydrogen densities $\log (n_{\rm H})$ from 5 to 12 cm$^{-3}$ (blue) and hydrogen column densities $\log(N_{\rm H})$ from 22 to 25 cm$^{-2}$ (orange). Representative values are labeled along the model tracks. Solid curves show models with a covering fraction of $f_{\rm c}=1$, while the dashed curves show the corresponding predictions for $f_{\rm c}=0.8$. The red star and associated error bar are the measurement obtained from the X-Shooter spectrum of UGCA320-IMBH.}
\vspace{-4mm}
\end{figure}

\begin{figure}[H]
\begin{center}
 \includegraphics[width=0.9\textwidth]{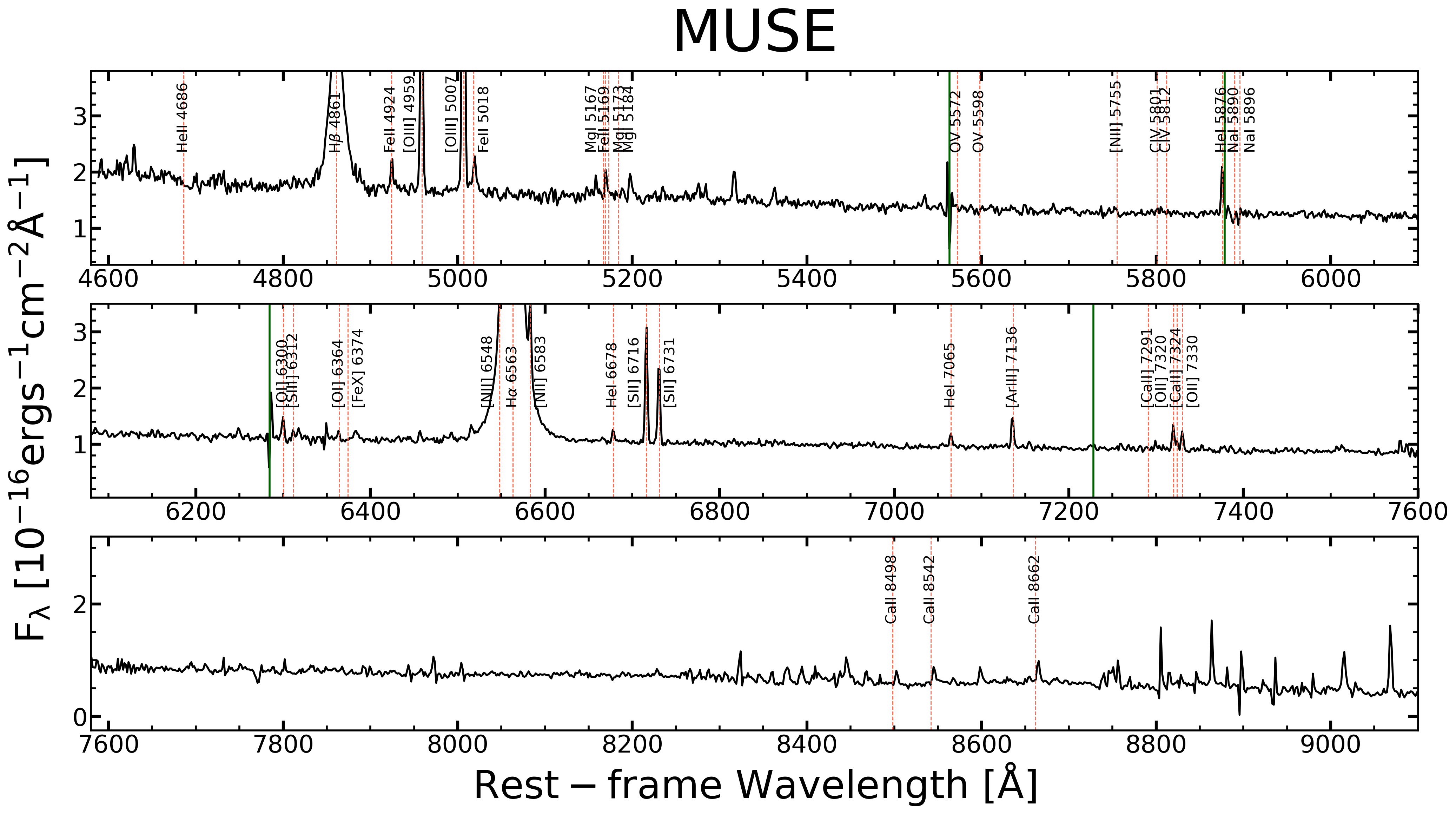}   
\end{center}
\vspace{-4mm}
\caption{\label{fig_muse} \textbf{MUSE stacked spectrum of the UGCA320-IMBH.} The spectrum is normalized to the median flux in each panel. Dashed red vertical lines mark the line centers, and the solid green lines mark the sky lines. }
\vspace{-4mm}
\end{figure}

\clearpage
\begin{addendum}
\item  [Acknowledgments]

L.X and Y.S. acknowledge support from the National Key R\&D Program of China (Nos.  2018YFA0404502  and  2017YFA0402704),  the  National  Natural Science  Foundation of  China (NSFC  grants 11825302,  11733002, and 11773013),  and the  Tencent Foundation  through the  XPLORER PRIZE. This work makes use of observations from the Las Cumbres Observatory global telescope network.

Based on data acquired at the ANU 2.3-metre telescope, under programes:2513051, 2613031. The automation of the telescope was made possible through an initial grant provided by the Centre of Gravitational Astrophysics and the Research school of Astronomy and Astrophysics at the Australian National University and through a grant provided by the Australian Research Council through LE230100063. The Lens proposal system is maintained by the AAO Research Data \& Software team as part of the Data Central Science Platform. We acknowledge the traditional custodians of the land on which the telescope stands, the Gamilaraay people, and pay our respects to elders past and present.

Based on observations collected at the European Southern Observatory under ESO programmes: 105.20GY.002, 115.29FU.001

Based on observations obtained with the Samuel Oschin Telescope 48-inch and the 60-inch Telescope at the Palomar Observatory as part of the Zwicky Transient Facility project. ZTF is supported by the National Science Foundation under Grants No. AST-1440341 and AST-2034437 and a collaboration including current partners Caltech, IPAC, the Oskar Klein Center at Stockholm University, the University of Maryland, University of California, Berkeley , the University of Wisconsin at Milwaukee, University of Warwick, Ruhr University, Cornell University, Northwestern University and Drexel University. Operations are conducted by COO, IPAC, and UW.

This work has made use of data from the European Space Agency (ESA) mission Gaia (https://www.cosmos.esa.int/gaia), processed by the Gaia Data Processing and Analysis Consortium (DPAC, https://www.cosmos.esa.int/web/gaia/dpac/consortium). Funding for the DPAC has been provided by national institutions, in particular the institutions participating in the Gaia Multilateral Agreement.

The Pan-STARRS1 Surveys (PS1) and the PS1 public science archive have been made possible through contributions by the Institute for Astronomy, the University of Hawaii, the Pan-STARRS Project Office, the Max-Planck Society and its participating institutes, the Max Planck Institute for Astronomy, Heidelberg and the Max Planck Institute for Extraterrestrial Physics, Garching, The Johns Hopkins University, Durham University, the University of Edinburgh, the Queen's University Belfast, the Harvard-Smithsonian Center for Astrophysics, the Las Cumbres Observatory Global Telescope Network Incorporated, the National Central University of Taiwan, the Space Telescope Science Institute, the National Aeronautics and Space Administration under Grant No. NNX08AR22G issued through the Planetary Science Division of the NASA Science Mission Directorate, the National Science Foundation Grant No. AST-1238877, the University of Maryland, Eotvos Lorand University (ELTE), the Los Alamos National Laboratory, and the Gordon and Betty Moore Foundation.

Based on observations obtained with MegaPrime/MegaCam, a joint project of CFHT and CEA/DAPNIA, at the Canada-France-Hawaii Telescope (CFHT) which is operated by the National Research Council (NRC) of Canada, the Institut National des Sciences de l'Univers of the Center National de la Recherche Scientifique (CNRS) of France, and the University of Hawai'i. The observations at the Canada-France-Hawaii Telescope were performed with care and respect from the summit of Maunakea, which is a significant cultural and historic site.

The scientific results reported in this article are based on observations made by the \textit{Chandra} X-ray Observatory, data obtained from the Chandra Data Archive.

This research has made use of data obtained from the \textit{Chandra} Data Archive and the Chandra Source Catalog, both provided by the Chandra X-ray Center (CXC).

This research has made use of data and/or software provided by the High Energy Astrophysics Science Archive Research Center (HEASARC), which is a service of the Astrophysics Science Division at NASA/GSFC.

\item[Author  Contributions] 

L.X. and Y.S. wrote the manuscript. Y.S. and F.B. led the Dwarf Galaxy Integral-field Survey, which observed the target galaxy, and oversaw the progress of this study. L.X. identified the target, led the follow-up X-shooter observations, and carried out the observational analysis. J.W. led the Chandra proposal. Y.J. performed the model to interpret the Balmer break. All authors discussed the results and commented on the manuscript.

\item[Data availability]  All data are available from the corresponding author upon reasonable request.

\item[Code availability] The codes used in this work from public packages. 

\item[Author  Information] Correspondence and requests for materials should be addressed to YS (shiyong@westlake.edu.cn), F.B. (fbian@eso.org) and J.W. (jfwang@xmu.edu.cn).

\item[Competing interests] The authors declare no competing interests.

\end{addendum}


\end{document}